\documentclass[article,11pt]{JHEP3}
\usepackage[iso-8859-7]{inputenc}
\usepackage{epsfig}
\usepackage{graphicx}

\usepackage{amsfonts}
\usepackage{amssymb}
\usepackage{amsthm}
\usepackage{amsmath}
\usepackage{multirow}

\usepackage{amssymb}

\let\e=\epsilon         
    \let\k=\kappa  \let\l=\lambda  
\let\n=\nu                 
        \let\f=\phi

  \let\D=\Delta

\def\nn{\nonumber}

\let\ov=\overline

\newcommand{\be}{\begin{equation}}
\newcommand{\ee}{\end{equation}}
\newcommand{\bea}{\begin{eqnarray}}
\newcommand{\eea}{\end{eqnarray}}
\newcommand{\ba}{\begin{array}}
\newcommand{\ea}{\end{array}}
\newcommand{\ra}{\rightarrow}

\usepackage{graphics}
\usepackage{epsfig}
\title{\Huge Phenomenological analysis of D-brane Pati-Salam vacua}

\author{\large
P. Anastasopoulos$^{1,2}$ \footnote{Pascal.Anastasopoulos@roma2.infn.it},~
G. K. Leontaris$^{2,3}$\footnote{leonta@uoi.gr},~
N.D. Vlachos$^4$\footnote{vlachos@physics.auth.gr},\\
$^1$ I.N.F.N.\ -\ Sezione di Roma ``Tor Vergata'', 00133, Roma, Italy\\
$^2$ Department of Physics, CERN Theory Division, CH-1211, Geneva 23, Switzerland\\
$^3$ Theoretical Physics Division, Ioannina University, GR-45110 Ioannina, Greece\\
$^4$ Theoretical Physics Division, Aristotle University, GR-54006 Thessaloniki, Greece}

\preprint{CERN-PH-TH/2010-026}

\abstract{
In the present work, we perform a  phenomenological analysis of the effective
low energy models with Pati-Salam (PS) gauge symmetry derived in the context of
D-branes.  The main issue in these models arises from the fact that the right-handed
fermions and the PS-symmetry breaking Higgs field transform identically under
the  symmetry, causing unnatural matter-Higgs mixing effects. We argue that this
problem can be solved in particular D-brane setups where these fields  arise
in different intersections. We further observe that whenever a large Higgs mass
term is being generated in a particular class of mass spectra,
a splitting mechanism --reminiscent of the doublet triplet splitting-- may protect
the neutral Higgs components from becoming heavy.  We analyze the implications of
each individual representation available in these models in order
to specify  the minimal spectrum required
to build up a consistent  model that reconciles the low energy data. A short discussion
is devoted to the effects of stringy instanton corrections, particularly those generating
missing Yukawa couplings and contributing to the fermion mass textures. We discuss
the correlations of the intersecting D-brane spectra with those obtained from
Gepner constructions and analyze the superpotential, the resulting mass textures and
the low energy implications of some examples of the latter along the lines
proposed above.
}

\begin{document}

\section{Introduction}

 Extended objects of the non-perturbative sector of string theory, the so-called
  D-branes \cite{bs, PS,Polchinski:1996na}, appears to be a promising framework for model
 building. Intersecting D-branes in particular, can provide chiral fermions
 and gauge symmetries, which contain the Standard Model spectrum and the
 $SU(3)\times SU(2)\times U(1)$ symmetry as a subgroup. The fermion and
 gauge fields are localized on the D-branes while gravity propagates in
 the bulk thus, D-brane models are natural candidates for phenomenological
explorations. During the last years, particular supersymmetric or
non-supersymmetric  models have been proposed, based on various D-brane
configurations, which exhibit a number of interesting properties.

Indeed, there are several remarkable features in these constructions that
convincingly point towards a thorough investigation
of promising D-brane derived models at low energy.  An interesting
property, for example, is that the multiplicity of the chiral sector
and the strength  of the Yukawa couplings are related to the geometry of the internal space.
 It has been also  shown that instanton contributions play a vital r\^ole in the
 interpretation of the hierarchical mass spectrum.
Furthermore, generically, the embedding of old successful Grand Unified Theories into
D-brane configurations enhance the old GUT gauge symmetries by
several  $U(1)$ factors, while some of them remain at low energies as global
symmetries. Interestingly, in certain occasions, there exist combinations of them,
which can be identified with  baryon or lepton conserving quantum numbers. At
the String level, these abelian symmetries contain anomalies, which are canceled
by a generalized Green-Schwarz mechanism. Usually, a linear
combination of these $U(1)$'s remains anomaly free and plays a
significant r\^ole in  phenomenological investigations~\cite{ak}-\cite{anto}.

In this work, we discuss in some detail D-brane models based on the Pati-Salam (PS) symmetry~\cite{Pati:1974yy}.
The realization of this gauge model is assumed in the context of type IIA intersecting D6-brane models.
However, the landscape of the string derived constructions is vast and it is not easy to select a viable
vacuum. Thus, to ascertain the required characteristics and incorporate the low energy phenomenological constraints of the string derived construction, in the first part of this paper, we will start our analysis with a bottom up approach. This will help us to pin down  the most promising models in Gepner constructions which will be discussed in the last sections.

As it is the case for all GUT models realized  within the context of intersecting D-branes,
the PS gauge group can also be accommodated  within a larger gauge symmetry, namely
$$U(4)\times U(2)_L\times U(2)_R  \cdot $$
Any gauge group factor  $U(n)$ of the above symmetry contains a decoupled $U(1)$ component,
and  we can locally write $U(n)= SU(n)\times U(1)$. Therefore, one ends up with an extended PS
symmetry accompanied by three $U(1)$ factors, which carry a strong impact on the superpotential
of the effective low energy theory.
Among other implications, they considerably restrict  the trilinear terms available in the superpotential, while in particular cases, a certain combination of them is  anomaly  free
 and can be used to modify the hypercharge generator. In several cases, such a modification
 could occur   without   affecting the hypercharges of the Standard Model content while
 it  might be used to provide integral charges to exotic states.

   In what follows, we will present the emerging massless spectrum
of the simpler D-brane configurations and determine the conditions under which
 realistic effective field theory models can be obtained.
Next, as a testing ground of our general analysis, we will attempt to work out some
semi-realistic examples derived from Gepner constructions where similar spectra arise.

In building up PS models from D-brane configurations, we are mainly faced with two problems.
One  arises from the fact that the right-handed fermions
and the PS breaking Higgs $H,\bar H$ transform exactly in the same way under the non-abelian part
of the gauge symmetry. Since Higgs representations usually appear in vector like pairs,
we are led to unacceptable  family and Higgs supermultiplets mixing. This mixing occurs via
effective mass terms generated from the trilinear part of the superpotential, after some appropriate singlet
fields develop non-zero vevs. A crucial observation in these models is
that these representations can appear under two different $U(1)$ charges, depending on whether
the relevant string-endpoint is attached to the $U(2)_R$ brane or its mirror.
We will subsequently show that this fact can be used to discriminate between the Higgs and
the right-handed fermions and avoid this way the Higgs-family mixing.
The second difficulty is closely related to the first one. It is quite frequent that
the representations accommodating the chiral matter are accompanied by anti-chiral
fields, and only the net number ($\#$ of chiral minus $\#$ of anti-chiral) can be identified
with the three  generations required. Consequently, there is an excess of vector like pairs,
which must become massive at a high ($\sim M_{GUT}$) scale and decouple from the low energy spectrum.
Whatever mechanism is mobilized to make these pairs massive, it will also provide
for a same order of magnitude   mass term  to the vector-like  Higgs fields  $H,\bar H$.
We will discuss the available mechanisms which provide sufficiently large masses to the extraneous
matter fields and, at the same time protect the Higgs field from receiving too large a mass.
As an alternative scenario, we will propose a mechanism (reminiscent to the doublet-triplet splitting)
which separates the charged particles from the neutral singlet, allowing the latter to develop a non-zero vev.
This vev will prove useful to generate masses for the various states through tree-level and
non-renormalizable Yukawa couplings.

The paper is organized as follows. In section \ref{PatiSalamFields},  we briefly describe the basic set up
of the $SU(4)\times SU(2)\times SU(2)$ (PS) gauge symmetry and discuss the minimal number of fields required to reproduce
the low energy Standard Model (SM) spectrum.  In section \ref{PatiSalamStrings} we present the general features of the
$U(4)\times U(2)\times U(2)$  D-brane analogue, paying particular attention
to the new features and their implications. In particular, we consider
 the implications of the additional $U(1)$ symmetries in detail (as in comparison to the minimal PS gauge group)  to
the form of the superpotential couplings. We further discuss the implications
of the  extraneous representations which accompany the Standard Model
spectrum. We present the  possibilities of accommodating the SM spectrum
in the D-brane intersections and exhibit the generic forms of the fermion
mass textures for each case. We discuss the neutrino sector and give
 a short presentation of the possible stringy instanton generated masses
whenever $U(1)$ symmetries or  anomaly cancellation restrictions do not
allow their explicit appearance in the perturbative superpotential.
In section \ref{PatiSalamHiggs} we present the Higgs sector and discuss  the possible
patterns of symmetry breaking down to the Standard Model gauge symmetry.
In section \ref{PSGepners} we discuss the
analogy between this generic D-brane spectra and comparable spectra derived from Gepner constructions,
and apply the above analysis to specific examples.  More details
together with further examples of  Gepner spectra are included in the appendices. Finally, in section
\ref{Concl} we present our conclusions.

\section{A Minimal Model with PS Symmetry}\label{PatiSalamFields}

In this section, we describe the minimal field theory version~\cite{Antoniadis:1988cm}
of the model based on the Pati Salam (PS) gauge symmetry
\bea
  SU(4)\times SU(2)_L \times SU(2)_R\label{PSGS} \cdot
\eea
This model has been extensively investigated in the context of the heterotic
superstring and it was shown to possess several attractive features.
Among them, we mention that this symmetry does not require the use of the adjoint
or larger Higgs representations to be broken down to the SM, while the doublet-triplet
splitting~\cite{Dimopoulos:1981zb} is not an issue here, since the color triplets and the Higgs doublets are no
longer members of the same multiplet. Furthermore, this model, in its minimal
version, predicts unification of the third generation Yukawa couplings while
the gauge couplings attain a common value at a scale as high as  $M_{GUT}\sim 2\times 10^{16}$GeV.

The matter field content of the minimal model consists of the following
representations. There are three chiral copies of $F_L$ and $\bar F_R$ multiplets transforming
 as the bifundamentals $(4,2,1)$ and $(\bar 4,1,2)$ respectively under the corresponding gauge
  symmetry factors in (\ref{PSGS}) which accommodate  SM fields including the right-handed neutrino.
  Both multiplets are integrated in the $16$ of the  $SO(10)$
  \bea
  16&\rightarrow&(4,2,1)+(\bar 4,1,2) \cdot
  \eea
  Employing the hypercharge definition
\bea
Y=\frac{1}{2}\,Q_{B-L} +\frac{1}{2}\, Q_{3R}
\eea
where
\bea
Q_{3R} =\left(
\begin{array}{cc}
 1 &0 \\
 0& -1
\end{array}
\right),&Q_{B-L}=\left(
\begin{array}{cccc}
 \frac 13 &0 & 0 &0\\
0 &\frac 13 & 0&0 \\
 0 & 0 & \frac 13&0\\
 0 & 0 &0& -1
\end{array}
\right)
\eea
the Standard particle assignment is
  \bea
F_L&=&({4},{2},{1})
=Q({3},{2},\frac 16) + \ell{({1},{2},-\frac 12)}\\
\bar{F}_R&=&(\bar{{4}},{1},{2})
=u^c({\bar 3},
{1},-\frac 23)+d^c({\bar3},{1},\frac 13)+
                           \nu^c({1},{1},0)+ e^c({1},
                           {1},1) \cdot \label{fermions}
\eea
    The $SU(4)\times SU(2)_R\rightarrow SU(3)_C\times U(1)_Y$
 breaking can be realized with two Higgs fields $\bar H=(\bar 4,1,2)$ and $ H=(4,1,2)$
 \bea
 \bar H=(\bar 4,1,2)&=&(u^c_H,\,d^c_h,\,e^c_H,\,\nu^c_H)
 \\
 H=( 4,1,2)&=&(\bar u^c_H,\,\bar d^c_h,\,\bar e^c_H,\,\bar \nu^c_H)
 \eea
 which descend from the $16$ and $\overline{16}$ of $SO(10)$ respectively
  \bea
  16_{\bar H}&\rightarrow&(4,2,1)+({\bar 4,1,2})\\
  \ov{16}_H&\rightarrow&(\bar 4,2,1)+({ 4,1,2}) \cdot
  \eea
  The particle assignment of $\bar H$ shares  the same quantum numbers with $\bar F_R$,
  whilst that of $H$ shares the conjugate. Both acquire vevs  along their sneutrino like components
  at a high scale
\bea
\langle H\rangle \,=\langle \bar\nu^c_H\rangle \,=\,M_{GUT},&\langle\bar H\rangle \,=\langle\nu^c_H\rangle \,=\,M_{GUT} \cdot
\eea
The SM symmetry  breaking is realized by means of a bidoublet field
$h=(1,2, 2)$. This bidoublet constitutes part of the  $10$ of $ SO(10)$
which, under the PS-chain breaking gives $10\rightarrow D_6(6,1,1)+h(1,2,2)$. After the spontaneous breaking
of the PS symmetry down to the SM, the sextets decompose to the usual triplet pair $D_6\ra D_3+\bar D_3$,
and the bidoublet to the two MSSM Higgs multiplets $h\ra h_u+h_d$ which subsequently
realize the SM breaking and provide masses to the fermions.

At the tree-level, all fermion species receive Dirac mass  from a common Yukawa term
$\bar F_R F_L h$. In the presence of a $U(1)$ family-like symmetry~\cite{Dent:2004dn} (as is the case
of heterotic string models for example), only the third generation receives tree-level masses,
and, at the GUT scale, Yukawa unification is predicted
\bea
\lambda_t(M_{GUT})=\lambda_b(M_{GUT})=\lambda_{\tau}(M_{GUT})=\lambda_{\nu}(M_{GUT}) \cdot
\eea
Furthermore, the PS symmetry implies the following mass relations at $M_{GUT}$
\bea
m_{t}(M_{GUT})&=&m_{\nu}(M_{GUT})\\
m_{b}(M_{GUT})&=&m_{\tau}(M_{GUT})
\eea
up to small threshold corrections. Smaller Yukawa contributions to the fermion masses arise from
non renormalizable terms involving the Higgs fields $H,\bar H$ and, eventually, a neutral
singlet vev $\phi$ charged under the $U(1)$ family symmetry.  Majorana masses which
realize the see-saw mechanism may arise from fourth order NR-operators
and -depending on the particular $U(1)$ charge assignment-
possible subleading terms of the form~\cite{Dent:2004dn}
\bea
M_{\nu^c}&\propto&\frac{\bar F_{Ri}\bar F_{Rj}HH}{M_{GUT}},\;\frac{\bar F_{Ri}\bar F_{Rj}HH\phi}{M_{GUT}},\;\;\cdots
\eea
In the presence of sextet fields $D_6$, the trilinear superpotential terms $HHD_6+\bar H\bar HD_6$
provide the triplets $d_H^c, \bar d_H^c$ (uneaten by the Higgs mechanism) with GUT-scale masses, by
pairing them up with the triplets of $D_6$, $M_{GUT}d_H^cD_3$ and $M_{GUT}\bar d_H^c\bar D_3$ as follows:
\bea
HHD_6+\bar H\bar HD_6\rightarrow M_{GUT}d_H^cD_3+M_{GUT}\bar d_H^c\bar D_3 \cdot
\eea

 Having exhibited the salient features of the PS model, we now turn on to the D-brane
  version, which in its minimal form is obtained from a three-stack D-brane configuration leading
  to the enlarged $U(4)\times U(2)\times U(2)$ gauge group.

\section{The D-brane Pati-Salam analogue}\label{PatiSalamStrings}

The D-brane realization of the PS symmetry requires a minimum of three stacks of
branes~\cite{Leontaris:1995sf}-\cite{Chatzistavrakidis:2010gz}.
Gauge fields are described by open strings with both endpoints attached on the same stack, and, they generically give
rise to Unitary groups\footnote{ Generically, D-brane stacks provide Unitary, $USp$ or $SO$ groups.   Replacing a U(2) with Sp(2) branes is always a possibility only if there is no contribution of the corresponding U(1) (within the U(2)) to the Hypercharge. On the other hand, models with Sp(2)'s contain vector-like reps and allow for more terms in the superpotential due to less constrains of gauge invariance. Notice further that the Pati-Salam gauge symmetry is isomorphic to $SO(6)\times SO(4)$, however, the spectrum and couplings would appear more constrained in the context of $SO$ groups. Therefore in this work, $USp$ and $SO$ groups can appear only in a hidden sector.}.
The rest of the SM particles corresponds to open strings attached on different (or the same) stack
providing bi-fundamental (as well as symmetric or antisymmetric) representations.
The hypercharge is a linear combination of the abelian factors of each stack. In general, the other linear
combinations of the abelian factors are anomalous. These anomalies are canceled by the Green-Schwarz mechanism and
 by generalized Chern-Simons terms~\cite{gcs}. The anomalous U(1) gauge bosons are massive and their masses can vary
 between the string scale and a much lower scale depending on the appropriate volume factors \cite{u1mass}.
The $F_L, ~ F_R,~F_L', ~ F_R',~H,~H'$ and their conjugates are described by strings with one end on the a-stack
(the 4 branes stack), and the other on the b- or c-stacks (the 2 branes stacks).
The $h,~h'$ and their conjugates correspond to strings stretched between the b- and c-stack.
Strings with one end on one stack and the other on the image stack, transform under
symmetric or antisymmetric representations.

Table \ref{U422} shows all the possible representations that can be generated by strings stretched
between the three brane-stacks and their corresponding mirrors. The appearance of any of these
states in the model spectrum  depends upon the particular choice  of the
intersections number which has to respect certain consistency conditions. In the intersecting
D-brane scenario, for example, one has to ensure appropriate numbers of intersections, which
lead to three chiral families and the disappearance of exotic matter.
At the same time, the solutions should respect the constraints ensuring  anomalies and tadpole
 cancelation~\cite{bs, PS, Aldazabal:2000dg}.

Although several D-brane variants have so far appeared in the literature~\cite{bu1}-\cite{Anastasopoulos:2006da},
a lot still remains to be done concerning a systematic investigation  of the resulting effective
low energy theory and its predictions. Up until now considerable work has been devoted to explore the implications
of the heterotic analogue. However, the issue of the D-brane construction carries a separate interest of its own.
The reason is that in comparison to the heterotic constructions, D-brane models contain new ingredients.
These ingredients predict new states not previously available, that include the adjoints and various symmetric representations under the three gauge group factors of the PS symmetry.
On the other hand, global $U(1)$ symmetries impose rather important restrictions on the superpotential couplings
at low energies, remnants  from the additional gauged
abelian factors discussed earlier. Thus, given the aforementioned differences with other (heterotic
string and orbifold) embeddings of the PS symmetry, the hope is that in working out a realistic brane model, the exotic matter
would give a clear signal that could discriminate it from other string constructions.
This way, we will consider a PS model carrying the general characteristics  of these D-brane variants
and discuss the implications, the viability and the prospects. In particular, we will attempt to specify
the minimal spectrum and the properties required for obtaining a viable low energy effective field theory.
In the next sections, we will also discuss the matching of
 the intersecting D-brane spectrum with spectra arising in certain Gepner  constructions and
 analyze a number of semi-realistic examples~\cite{Anastasopoulos:2006da}.

\begin{table}[!t]
\centering
\renewcommand{\arraystretch}{1.2}
\begin{tabular}{|c|c|ccc|}
\hline
Inters. & $SU(4)\times SU(2)_L\times SU(2)_R$& $Q_4$ & $Q_{2L}$ & $Q_{2R}$ \\
\hline
 $ab$   & $(n+k)\times F_L\,(4,\bar 2,1)$ & $1$ & $-1$ & $0$    \\
   & $n\times F_L\,(\bar 4, 2,1)$ & $-1$ & $1$ & $0$    \\
   \hline
 $ ab^*$   & $(m+k')\times F'_L\,(4, 2,1)$ & $1$ & $1$ & $0$    \\
   & $m\times \bar F'_L\,(\bar 4, \bar 2,1)$ & $-1$ & $-1$ & $0$    \\
   \hline
$ac$& $\ell\times \bar F_R\,(\bar 4,1,2)$ & $-1$ & $0$  & $1$ \\
& $\bar n\times \bar H\,(\bar 4,1,2)$ & $-1$ & $0$  & $1$ \\
& $\bar n\times  H\,(4,1,\bar 2)$ & $1$ & $0$  & $-1$ \\
\hline
$ac^*$& $\ell'\times \bar F'_R\,(\bar 4,1,\bar 2)$ & $-1$ & $0$  & $-1$ \\
& $\bar m\times \bar H'\,(\bar 4,1,\bar 2)$ & $-1$ & $0$  & $-1$ \\
& $\bar m\times  H'\,(4,1, 2)$ & $1$ & $0$  & $1$ \\
\hline
 $aa^*$   & $ D_6\,(6,1,1)$ & $\pm 2$ & $0$ & $0$        \\
          & $(S_{10},\bar S_{10})$&$\pm 2$&0&0\\
 $cc^*$   & $ \Delta_R\,(1,1,3)$ & $0$ & $0$ & $\pm 2$        \\
  & $  (\phi,\bar\phi)$&0&0&$\pm 2$\\
 $bb^*$  & $ \Delta_L\,(1,3,1)$ & $0$ & $\pm 2$  & $0$       \\
   & $(\nu_s,\bar\nu_s)$ & $0$ & $\pm 2$ & $0$        \\
\hline
 $bc^*$   & $\left.\begin{array}{c}h(1,2,2)\\
                 \bar{h}(1,\bar 2,\bar 2)
                 \end{array}\right.$  & $\left.\begin{array}{c}0\\
                 0
                 \end{array}\right.$ &$\left.\begin{array}{c}1\\
                 -1
                 \end{array}\right.$& $\left.\begin{array}{c}1\\
                 -1
                 \end{array}\right.$       \\
                 \hline
 $bc$   & $\left.\begin{array}{c}h'(1,2,\bar 2)\\
                 \bar{h}'(1,\bar 2,2)
                 \end{array}\right.$  & $\left.\begin{array}{c}0\\
                 0
                 \end{array}\right.$ &$\left.\begin{array}{c}1\\
                 -1
                 \end{array}\right.$& $\left.\begin{array}{c}-1\\
                 1
                 \end{array}\right.$       \\
\hline
\end{tabular}
 \caption{The Spectrum and the corresponding quantum numbers emerging in a D-brane configuration
with $U(4)\times U(2)\times U(2)$ gauge symmetry. $\ell,\ell',k,k',m,\bar m, n, \bar n$ represent multiplicities
of the corresponding states.
By $a,~b,~c$ we denote the stacks of 4-,~2$_L$- and 2$_R$-stack of branes respectively and ``*'' denotes the mirror branes under the orientifold planes. Multiple numbers of fields may also arise in the remaining intersections
$aa^*$, $bb^*$, $cc^*$, $bc$, $bc^*$.\label{U422}}
\end{table}

\subsection{The Superpotential }

We start our discussion with the superpotential of the effective theory which at the perturbative
level  may receive tree-level contributions and higher order corrections. It is to be noted that in these constructions
several tree-level superpotential terms of crucial importance are absent because of the surplus $U(1)$ symmetries.
In addition, in several cases non-renormalization theorems can also prevent the appearance of non-renormalizable terms.
 In this case, stringy instanton effects~\cite{Kiritsis:1999ss}-\cite{ACDL}\footnote{For recent reviews see
 \cite{Blumenhagen:2009qh}-\cite{Bianchi:2009ij}.} could compensate for the absence of
the missing terms.
This situation occurs usually in solutions with minimal spectra. In what follows, we make a detailed investigation
 of the perturbative and non-perturbative superpotential terms in order to pin down the minimal number of fields
 of Table \ref{U422} required to build a consistent model.

\subsubsection{Perturbative superpotential}

 We first observe that in the PS model fermions belong to bifundamentals
 created by strings whose endpoints are attached on the branes with $U(4)$   and $U(2)_{L/R}$ factors.
 Left-handed fermions transform as $4 \in SU(4)$ and are represented by strings attached on the $U(4)$ brane
 with $+1$ charge under the corresponding $U(1)$. Right-handed fermions transform as $\bar 4\in SU(4)$ and
 carry $-1$ charge under the same abelian factor. We have more options for the other endpoint of the string.
 Since in $SU(2)$  doublets and antidoublets are indistinguishable, we may choose to attach the
other endpoint of these strings either on
 the $U(2)$ branes or on their mirrors. Families attached to the $U(2)$ branes however, will differ from those attached
 to the mirrors with respect to the corresponding $U(1)_{L/R}$ factors. We may take advantage of this fact and make a suitable
arrangement
 of the families to obtain the desired fermion mass hierarchy and meet all the related requirements of  low-energy physics.
 We  assume  three families of left $F'_{L},F_{L}$ and right $ \bar F'_{R},\bar F_{R}$ fields, thus, the multiplicities
shown in Table must fulfill
\bea
 k+k'=\ell+\ell'=3.
\eea
We also assume   Higgs bidoublets $h,\bar h$
originating from the $bc^*$ intersection, corresponding to the two possible orientations of
the string~\footnote{The bidoublet representations $h$, and $\bar h$ arising in the intersection
$bc^*$ differ only under the two $U(1)$ charges. Since these $U(1)$'s do not participate in the
hypercharge, either of them  contain both $h_u, h_d$ doublets of the MSSM. Thus, in principle,
one of them could be adequate to realize the $SM$ symmetry breaking.}.
 Additional Higgs bidoublets (designated by $h',\bar h'$)  may also arise from the $bc$ intersection.

  The fermions of the three Standard Model families obtain their masses from gauge invariant couplings of
the form $\bar F_R^i\,F_L^j\,h$. However, due to the $U(1)$ symmetries, some of these terms might  not be allowed.
  Introducing indices for the representations with identical transformation properties, while assuming
  that the Higgs fields arise only in $bc^*$ intersection, (i.e., if  only $h,\bar h$  exist),
   the available tree-level couplings are
\bea
{\cal W}&=&\lambda_{im}\,\bar F'_{R_i}F_{L_m}\langle h\rangle+
\lambda'_{nj}\bar F_{R_n} F'_{L_j}\langle\bar h
\rangle \cdot \label{treemass}
\eea
If the  $h',\bar h'$  bidoublet Higgses are also present, we may  have the supplementary terms
\bea
{\cal W}'&=&
y_{nm}\bar F_{R_n} F_{L_m}\langle \bar h'
\rangle+y'_{ij}\,\bar F'_{R_i}F'_{L_j}\langle h'\rangle \cdot \label{treemass1}
\eea
In the minimal case where we have no-extra vector like pairs $F_L\bar F_L$ and $\bar F_RF_R$,
the  various fermion masses are obtained by the usual $3\times 3$  matrices whose formal structure
looks like
\bea
m_{u,d,\ell,\nu}&\sim&\left(
\begin{array}{ll}
 y_{nm}\langle \bar h'
\rangle & \lambda_{nj}' \langle\bar h
\rangle\\
 \lambda_{im}\langle h\rangle & y_{ij}'\langle h'\rangle
\end{array}
\right)
\eea
where the various indices $i,j,m,n$  take the appropriate values.

As we will soon see, the presence of both kinds of bidoublet Higgs fields ($ac$ and $ac^*$)
generates a number of undesired Yukawa couplings. Therefore, in a rather realistic construction,
we should be able to accommodate
only one kind of Higgs, (say $h$ and/or $\bar h$). In this case, we  distinguish between the following distinct
non-trivial classes of mass matrices.

\begin{itemize}

\item[A.]
If all the left handed representations arise from the $ab$ sector and the right-handed ones from
the $ac^*$ intersection, the only  surviving term is $\lambda_{im}\,\bar F'_{R_i}F_{L_m}\langle h\rangle$
and the mass matrices take the form
\bea
m_{u,d,\ell,\nu}&\sim&\left(
\begin{array}{lll}
\lambda_{11} & \lambda_{12} & \lambda_{13} \\
\lambda_{21} & \lambda_{22} & \lambda_{23} \\
\lambda_{31} & \lambda_{32} & \lambda_{33}
\end{array}
\right)\langle h\rangle \cdot \label{mat1}
\eea

\item[B.]
As a second possibility we consider the case where the left-handed fields arise from both
the $ab$ and $ab^*$ sectors, $F_{1L}',F_{2L},F_{3L}$, with all the right-handed ones descending from
the $ac^*$ intersection. Now, we get
\bea
m_{u,d,\ell,\nu}&\sim&\left(
\begin{array}{lll}
0 &0 &0\\
\lambda_{21} & \lambda_{22} & \lambda_{23} \\
\lambda_{31} & \lambda_{32} & \lambda_{33}
\end{array}
\right)\langle h\rangle \cdot \label{mat2}
\eea
In addition, two more cyclic permutations of the above
matrix can be obtained by interchanging the generation indices.  A comment is here in order.
The PS gauge symmetry implies in all cases the same texture form for
the up and down quarks. Higher order corrections are expected to discriminate between up and down
quark mass matrices and create the desired Cabibbo-Kobayashi-Maskawa (CKM) mixing.  We note that
this is in contrast to some
cases~\cite{Leontaris:2009ci}-\cite{Krippendorf:2010hj} which arise in the
context of SM gauge symmetry where the up and down quark mass matrices have a `complementary'
texture-zero structure making it hard to reconcile the CKM mixing~\cite{Leontaris:2009pi}.

\item[C.]
As a final possibility, we consider the case of $F_{1L}',F_{2L},F_{3L}$, and  $F_{1R},F_{2R}',F_{3R}'$
combination which leads to the following texture
\bea
m_{u,d,\ell,\nu}&\sim&\left(
\begin{array}{lll}
0 &\lambda_{12}\langle h\rangle &\lambda_{13}\langle h\rangle\\
\lambda'_{21}\langle\bar h\rangle &  0& 0 \\
\lambda'_{31}\langle \bar h\rangle & 0 & 0
\end{array}
\right) \cdot
\eea
In general, the above tree-level mass matrices contain several zeros which are expected
 to be filled  by contributions coming from non-renormalizable terms or
instantons.

\end{itemize}

The above textures deserve some discussion. In case A, we have seen that the
specific choice of field accommodation has led to a mass texture where all the entries
appear at the tree-level of the perturbative superpotential. In this case,
it is expected that all Yukawa couplings $\l_{ij}$ are  of the
same order of magnitude. In such a texture, the hierarchical
pattern will be rather difficult to  explain in a natural way.
Case B leads to a texture zero mass matrix with
three zeros in the first row. Due to the PS symmetry, there is a unique
gauge invariant Yukawa coupling to account for the fermion mass matrices,
thus, we end up with the same texture for all kinds of fermions, namely
up,  down, charged leptons and Dirac neutrinos. Since non-perturbative
contributions and/or NR-terms are expected to fill the gaps, and these contributions
 are naturally suppressed as compared to tree-level terms, this texture  looks more realistic
than the case A. Finally, the structure of case C is rather peculiar and probably
less suitable for the charged fermion mass spectrum.

Let us now deal with the $SU(4)$ breaking Higgs fields. These may arise at
the $ac$ and $ac^*$ intersections and are denoted by $H+\bar H$, and $H'+\bar H'$
respectively. Since  $\bar H,\bar H'$ carry exactly the same quantum numbers as
$\bar F_{R_i}, \bar F'_{R_i}$ correspondingly,
we must also have the couplings
\bea
{\cal W}''&=&\lambda_{ij}\,\bar H'\,F_{L_j}\, h+
\lambda'_{ij}\bar H\, F'_{L_i}\,\bar h \cdot
\eea

In order to break the $SU(4)\times SU(2)_R$ symmetry, we have to assign  vevs either to  $H+\bar H$, or to $H'+\bar H'$.
 Allowing  the RH fields to descend from both $ac$ and $ac^*$ intersections, anyone of the  $\bar H, \bar H'$
vevs will render at least one lepton and one Higgs doublet massive at an
unacceptably large scale
\bea
\lambda_{ij}\,\langle\bar H'\rangle\,F_{L_j}\, h&\rightarrow& M_{GUT} \ell_j h_u\label{fatal}\\
\lambda_{ij}\,\langle\bar H\rangle\,F'_{L_j}\, \bar h&\rightarrow& M_{GUT} \ell'_j \bar h_u\nn \cdot
\eea
 This problem may be evaded by assuming that the $SU(4)$ breaking Higgs and the representation $\bar F_R$
accommodating the right-handed fields have distinct quantum numbers under $U(1)$ symmetries.
This is possible under the following arrangement:

First, we will assume that all three representations $(4,2,1)$ accommodating the left handed
fields arise in the $ab$ intersection, thus $k=3,k'=0$. We will further demand
that the right-handed fields are only at the $bc^*$ intersection thus $\ell=0, \ell'=3$. Then,
only the first tree-level coupling in (\ref{treemass}) is present at ${\cal W}$:
\bea
\lambda_{ij}\,\bar F'_{R_i}F_{L_j}\,h&\rightarrow & 
Q\,u^c\,\langle h_u\rangle +\ell\,\nu^c\,\langle h_u\rangle
+ Q\,d^c\,\langle h_d\rangle +\ell\,e^c\,\langle h_d\rangle \cdot
\eea

 Next, in order to avoid undesired similar couplings, we will impose $\bar m=0$, and $\bar n\ne 0$ (preferably $\bar n=1$),
 so we have only  $H+\bar H$ pairs, which  carry different $U(1)$
 charges from $\bar F'_{R_i}$'s. This choice prevents the appearance of the unwanted couplings (\ref{fatal}).

The $SU(4)$ breaking Higgs fields $H+\bar H$ involve $\bar u_H^c, u_H^c, \bar e_H^c, e_H^c$
`eaten' by the Higgs mechanism, and `uneaten' down-quark type color triplets
$\bar d_H^c, d_H^c$  which must become massive at a high scale.
In the presence of the sextet fields $D_6 =D_3+ D^c_3$ and $\bar D_6 =\bar D_3+\bar D^c_3$,
the simplest way to realize this is via  couplings of the form
$HH\bar D_6, \bar H\bar H D_6$, which in the intersecting D-brane scenarios are not allowed by the  $U(1)$ symmetries
at the tree-level. Fortunately, there are alternative ways to obtain masses for the triplets in these constructions.
These involve the inclusion of non-renormalizable terms in the presence of additional singlet fields, which develop appropriate
vevs and/or, possible instanton effects.

 We start with the first approach. We observe that the  singlet fields  $\phi,\bar\phi$  arising from the $cc^*$
 sector and  presented in Table \ref{U422}, have the appropriate $U(1)$ charges to generate the fourth order terms
\bea
\frac{\phi}{M_S}\,HH\,\bar D_6,\;\; \frac{\bar\phi}{M_S}\,\bar H\bar H D_6 \cdot \label{Dphiterms}
\eea
 Upon developing vevs $\langle\phi\rangle\sim\langle\bar\phi\rangle\sim M_{GUT}$, the singlet fields generate
 the missing mass terms for the triplets. We should point out however, that  possible non-zero vevs for
 these singlets, would generate  undesired effects as it is the case, for example, for the term $\bar F_{R}\, \phi\, H$.
In such cases, we are forced to set $\langle\phi\rangle=0$ nevertheless we will see that
 it is possible to derive the mass terms (\ref{Dphiterms}) by instanton effects.

To determine the final form of the down-type colored triplet sector, one should also encounter the
terms $\bar H\,\bar F'_{R}\,D_6$ and $H\,F_L\,\bar D_6\,h$ which mix the down quarks in $\bar F'_R$
with the triplet fields living in $D_6,\,\bar D_6$, leading thus to an extended down quark mass matrix.
Higher order non-renomalizable terms may also contribute to the generalized down quark mass matrix,
which, under a judicious choice of the various field vevs for $\phi, H,\bar H$
can leave three light eigenstates to be identified with the ordinary down quarks.

Proceeding with the analysis, we recall that as opposed to the minimal field theory version
presented previously, in  D-brane constructions
there exist additional states, which belong to the symmetric and/or antisymmetric representations of
each non-abelian gauge group factor.  These are designated by  $S_{10},\bar S_{10}$ for
the $SU(3)$ case and $\Delta_{L,R}$ for the $SU(2)_{L,R}$ respectively\footnote{In the presence
of bulk branes, additional states in  $(4/\bar 4,1,1)$, $(1,2,1)$ and $(1,1,2)$ are also
possible}. The additional $SU(2)_R$ triplets   have the
particle assignment $\Delta_R(1,1,3)=(\delta^+,\delta^0,\delta^{-})$, or
\bea
\tau\cdot{\Delta}_R&=&\left(
\begin{array}{cc}
 \frac{\delta^0}{\sqrt{2}} &\delta^+ \\
 \delta^{-}&  -\frac{\delta^0}{\sqrt{2}}
\end{array}\label{3R}
\right)
\eea
and analogously for the  $\Delta_L(1,3,1)$ ones. Also, the $S_{10}$ SM-decomposition involves
\bea
 S_{10}&\rightarrow&S_6(6,-\frac 23)+S_3( 3,\frac 23)+S_1(1,2)\label{S10}
 \eea
where the triplet field $S_3$ carries  the quantum numbers of the down quarks.
We will see that in  cases where more conventional representations (i.e., the
neutral singlets and color sextet fields $D_6$)  are absent from the massless spectrum,
the fields (\ref{3R}) and (\ref{S10}) provide supplementary terms in the superpotential
which can act as surrogates to make  the color triplets in $H,\bar H$ Higgs fields massive,
or even realize the see-saw mechanism.

 Indeed, in models containing the representations $S_{10}$ and $\bar S_{10}$
 we may also have the trilinear couplings
\bea
H\,H\,\bar S_{10}+\bar H\,\bar H\,S_{10}&\rightarrow&M_{GUT}\,(\bar d_H^c\,\bar S_3+d_H^c\,S_3) \cdot \label{MS3}
\eea
The fields $S_{10}$ and $\bar S_{10}$, under the Standard Model gauge group decomposition involve also other exotic
representations, which should become massive at a high scale. This can be realized by a mass term $M_{10}S_{10}\bar S_{10}$
generated by a superpotential term where an effective scalar component $\phi_0$ acquires a non-zero vev
 at a scale  $\langle\phi_0\rangle \sim M_{GUT}$\footnote{If we restrict to the representations of Table 1, we can think
of such a scalar vev as the condensation of $\langle\phi\bar\phi\rangle$ where both singlets develop equal vevs.
For reasons that will become clear later, we require $\langle\phi_0\rangle \sim M_{GUT}$. This can be achieved
naturally assuming $\langle\phi\rangle\le 10^{-1/2}M_S$, and $M_{GUT}\le 10^{-1}M_S$, where $M_S$ is assumed
to be the String scale. Alternatively, $\langle\phi_0\rangle$
could be a vev of  the $U(4)$ adjoint. }.
In this case, a term $M_{GUT}\bar S_3S_3$ is also implied which, in conjunction with the mass terms (\ref{MS3}) form a triplet mass matrix with eigenmasses being naturally of the order $M_{GUT}$.

Finally, we also comment on the existence of an additional Higgs pair $h'+\bar h'$ in the intersection $bc$.
Denoting $h'=h_u'+h_d'$ and $F_L=Q+\ell$, we observe that the coupling
  \bea
 \bar H F_L  h'&\rightarrow & \langle\nu_H\rangle \,\ell\,h_u'
  \eea
 couples the lepton doublet $\ell$  with $h_u'$ through a mass of the order of  GUT scale.
 We could be further elaborate on this case  by constructing the full doublet mass matrix,
 taking into account possible non-renormalizable terms, and seek  solutions with three light
 doublets in analogy to the down quark mass matrix discussed above. This line, however,
 would lead to a rather contrived model, thus, it is more natural to assume the simpler
 case with only one light bidoublet Higgs $h$ and/or $\bar h$.

\subsubsection{Neutrino masses}

In PS models right-handed neutrinos are incorporated into the same representation with the
 right-handed charged fermions. They receive Dirac masses through the same term
(\ref{treemass}) sharing the same Higgs doublet with the up-quarks. Therefore, the Dirac
neutrino masses  are of the same order of magnitude with the up quark masses and a see-saw mechanism is
required in order to generate light Majorana mass eigenstates compatible with the present experimental
bounds.

In the presence of the fields $\Delta_R$ and $\phi$, an extended see-saw mechanism can generate light
 Majorana masses through the following tree level terms
\bea
\lambda'_{\nu^c}\bar F'_{R_i}\Delta_R\,H+\lambda_{\nu^c}\bar F'_{R}\, \phi\, H
&\to&\langle \tilde{\bar\nu}_H^c\rangle \delta^0{\nu'}^c+\langle \tilde{\bar\nu}_H^c\rangle\, \phi\,{\nu'}^c
\eea
with $\delta^0\in\Delta_R$. Either one of these terms is sufficient to realize the see-saw mechanism.
We also note that in the presence of $\bar\phi$ singlets, non-renormalizable mass terms contributing
to the neutrino mass matrix are also possible
\bea
\frac{\bar F'_{R}\bar F'_{R}HH\bar\phi}{M_{GUT}}&\sim&M_{nr}{\nu'}^c{\nu'}^c \cdot
\eea
Suppressing generation indices, the complete tree-level neutrino mass matrix in the basis $\nu_i,\nu_i^c$,
and $\delta^0$ (and/or $\phi$) can be written as
\bea
M_{\nu}&\sim&\left(
\begin{array}{lll}
 0 & m_u & 0\\
 m_u& M_{nr} & M  \\
 0 & M & 0
\end{array}
\right) \cdot \label{xtd}
\eea
Thus, taking into account all the contributions to  neutrinos and other neutral states,
we end up with  the extended see-saw type mass matrix (\ref{xtd}) which leads to three
light left-handed neutrino states, which can naturally lie in the sub-eV range
as required by the present neutrino data.

\subsubsection{ Instanton induced masses}\label{Instantsection}

We have seen in the previous sections, that for certain cases of D-brane spectra,
 several  Yukawa couplings of crucial importance are absent from the tree level superpotential.
 For  example, in the absence of the $h'+\bar h'$ bidoublets we noticed that Yukawas implying
  $\bar F_RF_L$ and $\bar F_R'F_L'$ mixings are not possible.
Similarly, the terms $HHD_6$ and $\bar H\bar H D_6$ are prevented by global $U(1)$ symmetries
leaving  the dangerous color triplets  \footnote{If $H',\bar H'$ were present, these
could have the following  perturbative Yukawa couplings
$\lambda_H\,H\,H'D_6 +\lambda_{\bar H}\,\bar H\,\bar H'D_6$.} massless.  Furthermore, in the absence
of appropriate singlet scalar fields with non-vanishing vevs, non-renormalizable contributions
are impossible.

 It has been suggested~\cite{Blumenhagen:2006xt, Ibanez:2006da, Florea:2006si}
 that  for a matter fields operator $\prod_j\Phi_{a_jb_j}$ violating the
 $U(1)$ symmetry, it is possible that Euclidean D2 (${\cal E}2$ for short)  instantons
having the appropriate number of intersections with the $D6$-branes can induce
non-perturbative superpotential terms of the form
\bea
{\cal W}_{n.p.}&\supset&\prod_{j=1}^J\Phi_{a_jb_j}\,e^{-S_{\cal E}} \cdot \label{instcoup}
\eea
In other words, the perturbatively forbidden Yukawa coupling is now realized
non perturbatively, since in the presence of appropriate instanton zero modes,
 the instanton action $S_{\cal E}$ can absorb the $U(1)_a$ charge
excess of the field operator violating the $U(1)$ symmetry. Indeed,
 under the  $U(1)_a$  symmetry the transformation property of the exponential
instanton action is
\bea
\,e^{-S_{\cal E}}&\rightarrow&\,e^{-S_{\cal E}}\,e^{\imath\,{\cal Q}_a({\cal E}2)\Lambda_a}\label{npc}
\eea
where ${\cal Q}_a({\cal E}2)$ represents the amount of the $U(1)_a$-charge violation
by the ${\cal E}2$ instanton.
If $\pi_a,\pi_{a^*}$ are the homological three-cycles of the $D6_a$ brane-stack and its
mirror respectively, then this is given by
\bea
{\cal Q}_a({\cal E}2)&=&-{\cal N}_a\,\pi_{\cal E}\circ (\pi_a-\pi_{a^*})\equiv -{\cal N}_a\,
(I_{{\cal E}a}-I_{{\cal E}{a^*}})\label{instcha}
\eea
where the $I_{{\cal E}a}$ and  $I_{{\cal E}{a^*}}$ stand for the relevant intersection numbers.
For rigid $O(1)$ instantons, wrapping a rigid
orientifold-invariant cycle in the internal space,  due to the ${\cal E}2-a$ and
${\cal E}2-a^*$ identification the charge (\ref{instcha}) simplifies to
\bea
{\cal Q}_a({\cal E}2)&=&-{\cal N}_a\,\pi_{\cal E}\circ \pi_a\equiv -{\cal N}_a\,
I_{{\cal E}a}\label{E2D6}
\eea
Thus, allowing for an appropriate number of wrappings, the above can exactly match the
$U(1)_a$ charge excess of the filed operator $\prod_j\Phi_{a_jb_j}$ and the total
coupling (\ref{npc}) is $U(1)_a$-invariant.
\begin{figure}
\begin{center}
\epsfig{file=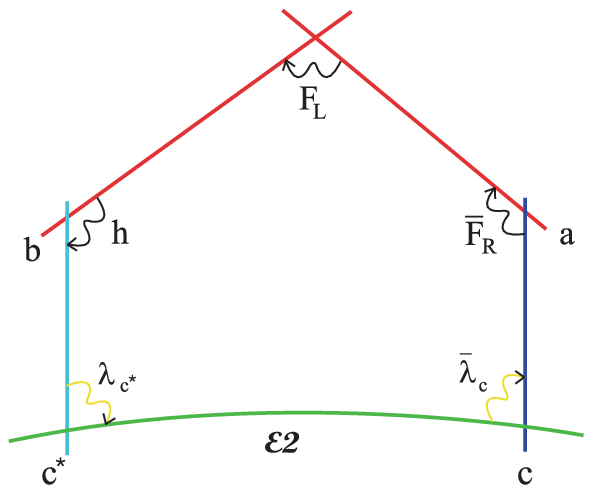 ,width=50mm}$\;\;\;$
\epsfig{file=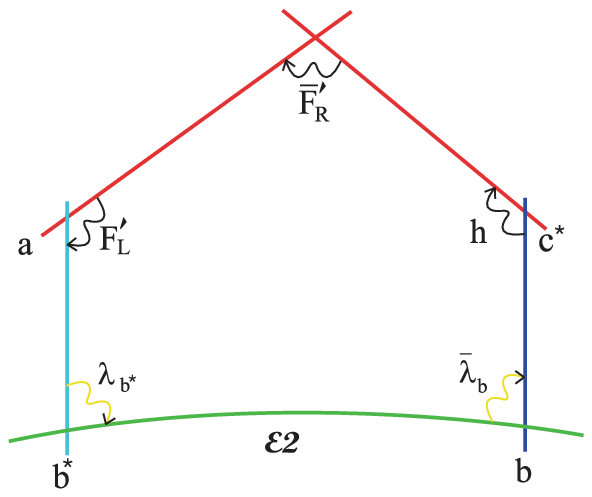 ,width=50mm}
\end{center}
\caption{Stringy instanton generated  Yukawa coupling $F_RF_Lh,\,F_R'\,F'_L\,h$.}\label{flr}
\end{figure}

Returning to our specific model discussed here, we can, for example, observe  that when $h',\bar h'$
bidoublets related to the intersection $bc$, are not found in the massless spectrum, the couplings
(\ref{treemass1}) are not available.  It is then possible  that the zero entries in
the fermion mass matrices discussed in the previous section, are filled  by the instanton
contributions (see fig \ref{flr})
\bea
{\cal W}_{n.p.}&=&\lambda_1^{np}\,F_RF_Lh+\lambda_2^{np}\,F_R'\,F'_L\,h \cdot \label{fnpc}
\eea
The induced coupling (\ref{instcoup}) involves an exponential suppression by the classical instanton
action ${\cal W}_{n.p.}\propto \exp\{-\frac{8\pi^2{\rm Vol}_{\cal E}}{g_a^2{\rm Vol}_{D6_a}}\}$.
This way, the couplings $\lambda_{1,2}^{np}$ are suppressed by exponential factors involving the
classical instanton action, and are expected to be much smaller than  the perturbative
ones.
\bea
\lambda_{i}^{n.p.}&\sim &{\cal O}\left(e^{-S_{\cal E}}\lambda_i\right) \cdot
\eea
Note that other instantons can generate  additional contributions to the same
fermion zero entries,   inducing factorizable Yukawa couplings instead of (\ref{fnpc}).
This latter type of couplings appears also for the sextets fields. In particular,
the zero mode wrapping conditions  $I_{{\cal E}a}=1$  and $I_{{\cal E}b}=-1$ allow for the
coupling $HHD_6$. Similar instanton effects with the appropriate winding numbers may generate
couplings of the form $\bar H\bar H D_6$.
We should remark however, that when both terms are present, potentially dangerous
 dimension five proton decay operators are generated as it was also observed in~\cite{Cvetic:2009mt, Anastasopoulos:2009mr, Kiritsis:2009sf}. A crucial r\^ole is then played by the the scale
at which the triplets $D_3,\bar D_3\subset D_6$ become massive.\footnote{ For proton decay  issues
  in the corresponding field theory models see for example~\cite{Babu:1997js}.}
\begin{figure}
\begin{center}
\epsfig{file=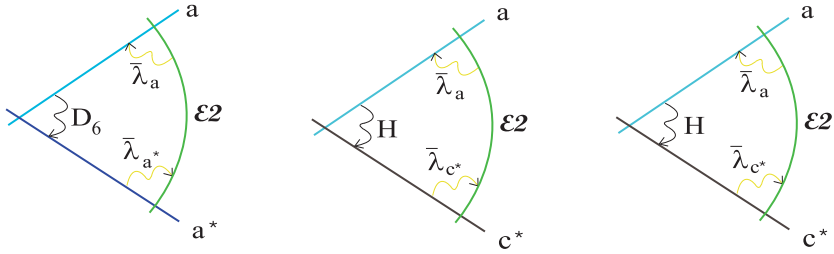 ,width=110mm}
\end{center}
\caption{Stringy instanton generated  Yukawa coupling $HHD_6$ giving mass to the color triplets.}\label{D6HH}
\end{figure}

\section{The Higgs Sector and the  right-handed ``Doublet-Triplet'' splitting}\label{PatiSalamHiggs}

We have already pointed out that a rather general phenomenon in these constructions
is the appearance of extra matter fields beyond those contained in the Standard Model.
Our interest in the present section will focus on the extraneous matter  related
to the Higgs problem mentioned in the introduction. This particular extra matter comes
in vector like pairs. Such a case was  already encountered in the previous section, namely the $SU(3)$ non-trivial
states $S_{10}+\bar S_{10}$. The most common case consists of the representations
accommodating  the same chiral fields themselves.  Indeed, as it is generally shown in the Table,
one might have $n+3$ chiral states in $F_L$ and $n$ states in $\bar F_L$, thus
the net number of chiral supermultiplets is three;  however, there are $n\ne 0$ copies
of vector-like representations $F_L+\bar F_L$ since  it is rather difficult in practice to derive
a D-brane spectrum with $n=0$.  The same is also true for the right handed partners  $F_R, \bar F_R$
which in addition to the three chirals $\bar F_R$ may also have an excess of
several $\bar F_R+F_R$ copies.  In a viable effective field theory model, all these states should decouple at a high
scale. A natural mechanism to deal with this situation is to allow a non-zero vev to a singlet
field that couples to these states $\langle \phi_0\rangle (\bar F_LF_L+\bar F_RF_R+\dots) $, $ \langle \phi_0\rangle \sim M_{GUT}$, in order to avoid large threshold effects in the renormalization group (RG) running of precisely measured quantities at $M_W$ (i.e., $\sin^2\theta_w,a_{em},a_s$ etc).
In this case, it is unavoidable that a mass term $M_{GUT}H\bar H$ is generated,
implying that the $H,\bar H$ fields decouple from the spectrum and cannot develop vevs along their neutral components.
Another complication related  to the same Higgs fields emerges from the fact that  $\bar H$ and $\bar F_R$
transform identically under the PS symmetry, sharing thus the same Yukawa couplings and leading
to the unacceptable matter-Higgs mixing already discussed in the previous sections.
In what follows, we will suggest a  solution to avoid these problems.

We start with the second issue.
Taking the previous analysis at face value, we infer  that in order to avoid undesired
mixings, the RH fields $\bar F_R$ should not share the same  Yukawa couplings with the
Higgs $\bar H$. Since both of them transform equivalently under the non-abelian part of the
gauge symmetry, they can only differ with respect to the $U(1)_R$ gauge factor. Thus, we may choose
all RH-fermions from the $ac^*$ sector and pick up the $SU(4)$ breaking Higgs pair from the $ac$ intersection.
This arrangement  solves the problem of large unacceptable $H\bar F_R$ mixing. However, both $H,\bar H$ Higgs
come from the same intersection with opposite $U(1)$ charges as indicated in the Table.
If no extra matter in vector-like form is present in the spectrum of the theory, then there is
no need for developing non-zero singlet vevs which could couple to the Higgs pair $\bar HH$.
Consequently, the Higgs fields remain in the massless spectrum of the theory and their r\^ole
as discussed in the previous sections remains intact.

We have argued that quite often vector-like fields are unavoidable, while
consistency with  low energy data requires that these should decouple from the
light spectrum at a high scale. The most natural candidate mass terms for the extra pairs may
arise at fourth order and can be of the form
\bea
\frac{\langle\bar HH\rangle}{M_S}\left(\bar F_LF_L+\bar F_RF_R +\bar S_{10}S_{10} +\bar HH+\bar hh+\cdots\right) \cdot
\label{bili}
\eea
Assuming vevs of the Higgs fields along the D-flat direction,
we  naturally expect them to be of the order $\langle H\rangle \sim \langle\bar H\rangle \sim 10^{-1} M_S$
 leading to an adequately large mass  $M\sim 10^{-1} M_{GUT}$ for all extra matter.
Unfortunately, a large number of such fields would generate sizable threshold effects in the renormalization
group running of the gauge couplings,  spoiling this way the unification prediction.

As an alternative to the NR-contributions we can let a singlet develop a non-zero vev
$\langle\phi_0\rangle\sim M_{GUT}$
but as  already discussed above, this would lead into the problem of a similar Higgs
mass term.
Below, we present a mechanism that could resolve this problem.
In this case, whenever a mass term for  $\bar F_LF_L$, $\bar F_RF_R$
is generated, the $U(1)$ symmetries cannot prevent the appearance of the analogous  mass term for the Higgs pair
$M_H\,H\bar H$, where $M_H$ is naturally of the order of the Unification scale $M_{GUT}$. Note
that in the presence of an $SU(4)$ adjoint  Higgs field $\Phi_4$, the following terms are allowed;
\bea
{\cal W}_H\supset \bar H\,\Phi_4\,H+M_H\,\bar H\,H \cdot \label{4H}
\eea
 We may  now solve the problem by allowing  the  $SU(4)$-adjoint scalar $\Phi_4$ to obtain
a vacuum expectation value which cancels the $M_H$ term.  The $SU(4)$ adjoint  is
\bea
(15,1,1)&=&\left(\begin{array}{cc}
A_8-\frac{A_0}{2\sqrt{3}}\times {\bf 1}_3&B\\
\bar B&\frac{\sqrt{3}}{2}A_0
\end{array}
\right) \label{1511}\cdot
\eea
On the left-hand side of the above we explicitly indicate the transformation
of the adjoint representation under
the $SU(4)\times SU(2)_L\times SU(2)_R$ gauge symmetry.
Take  $v=-\langle \frac{\sqrt{3}A_0}{2}\rangle$, so that
\bea
\langle\Phi_4\rangle &=&\left(
\begin{array}{cccc}
\frac v3 &0 & 0 &0\\
0 &\frac v3 & 0&0 \\
0 & 0 & \frac v3&0\\
0 & 0 &0& -v
\end{array}
\right) \cdot
\eea
The resulting mass terms from (\ref{4H}) are written as
\bea
{\cal W}_H\supset\left(M_H-{v}\right)\,{\bar L}_H^c\,L_H^c+
\left(\frac{v}3+M_H\right)\bar Q^c_H\,Q^c_H  \label{4HA}
\eea
where $Q^c_H=(u_H^c,d^c_H)^T$, $L^c_H=(e_H^c,\nu^c_H)^T$ are $SU(2)_R$ doublets. Clearly,
because of the trivial transformation properties of the $SU(4)$ adjoint under the $SU(2)_{L/R}$
gauge groups (see (\ref{1511}), left-hand side), the corresponding gauge group factors are preserved.
The choice $v\sim M_H$ would leave the right-handed doublets ${\bar L}^c_ H,\,L_H^c$ of  $H,\bar H$ massless,
while at the same time  would supply all additional color particles with heavy masses $M\sim \frac{4}{3} M_H$
and break the original symmetry  down to a left-right symmetric model
\[
SU(3)\times SU(2)_L\times SU(2)_R\times U(1)_{B-L} \cdot
\]
 Thus, in this scenario, the PS symmetry breaking can occur in two  steps. The vev $\langle\Phi_4\rangle$
realizes the $SU(4)$ symmetry breaking while, the vevs  of the RH doublets ${\bar L}_H^c,\,L_H^c$
along their neutral directions  can break the $SU(2)_R$ symmetry. Although, in principle, these two
breaking scales could differ substantially, purely phenomenological requirements demand that the
$SU(2)_R$ breaking scale should not be much lower than the $SU(4)$ scale. Indeed,
this arrangement permits several tree-level and NR-terms to generate useful mass contributions for the
various SM fields, when $H,\bar H$ fields are allowed to acquire vevs along the neutral directions.
If the $SU(2)_R$ breaking scale is substantially smaller than $M_{GUT}$, these contributions would be
suppressed and therefore, become irrelevant.  Since the natural expansion parameter in NR contributions
is $\e\sim \frac{\langle\bar HH\rangle}{M_S}$,  we may    assume  that
$\langle H\rangle \sim \langle \bar H\rangle\sim v\sim M_{GUT}$, thus $\e$ takes natural values  $\e\sim{\cal O}( 10^{-1})$.  {This estimate is quite reasonable and is not in conflict with the Renormalization
Group analysis of the field theory model. Indeed, for mass spectra not deviating considerably from the minimal one, renormalization group analysis shows~\cite{Dent:2006nm} that the magnitude of these $SU(4)$-breaking  vevs cannot differ substantially from the standard SUSY GUT scale $ M_{GUT}\sim (2-3)\times 10^{16}$ GeV. Taking into account that $M_{String}\approx 2\times 10^{17}$ GeV, we conclude that  $M_U/M_{S}\sim 10^{-1}$. On the other hand, the various $M_{GUT}$ vevs including those of the $SU(4)$ breaking Higgs fieds $\langle\bar H,H\rangle$ are related to the FI terms and their values depend on the various moduli\footnote{More precisely, there is no 1-loop FI term in orientifold models \cite{poppitz} due to tadpole cancelation. The only contribution to FI term is coming at tree-level by the distance between the D-branes and the orientifold planes.}.
The solutions for particular Gepner models to be discussed later on show consistency with the above estimates.}

From the last term in (\ref{bili}) we observe that in general there exist analogous effective
mass terms for the light Higgs bidoublet fields as well. These couplings  remove  pairs $h_i,\bar h_i$
of left-handed doublets  from the light spectrum.  To implement the $SU(2)_L$ symmetry
breaking and provide fermions with masses, we need at least the content of one bidoublet Higgs $h\ra h_u+h_d$
in the massless spectrum.  Since bidoublet fields include both electroweak Higgs
doublets, they are not necessarily required to appear in pairs $h,\bar h$. So,
by arranging that  the number $n_h$ of $h$'s does not coincide with the number $n_{\bar h}$,
 of $\bar h$ fields,  a number of $|n_h-n_{\bar h}|$ bidoublets could, in principle,
 remain massless. As it will become evident in
 the next sections, this is exactly the case for the several examples obtained in the Gepner constructions. Also,
  in the case of $n_h=n_{\bar h}$ a splitting mechanism analogous to the $SU(4)$ case
 described above could be activated for the bidoublets where now the r\^ole of $\Phi_4$
 is played by the vev of an  $SU(2)_R$ triplet $\Delta_R$ or the
adjoint $\Phi_{2R}$~\footnote{More precisely, the adjoint could be used to realize the splitting
among bidoublets of the same intersection ($ac$ or $ac^*$), whilst
the triplets carry $U(1)$ charge and could be used to create a splitting
for mass terms ``mixing'' $ac$ and $ac^*$  bidoublets.}. Implementing this mechanism, while
assuming $\langle \Phi_{2R}\rangle ={\rm diag}(v_R,-v_R)$, we  obtain
 \bea
{\cal W}_h&\supset& \left(\frac{\langle H\bar H\rangle}{M_S}-v_R\right) h_u\bar h_d+
  \left(\frac{\langle H\bar H\rangle}{M_S}+v_R\right) \bar h_u h_d
\eea
with $\langle H\bar H\rangle\sim M_{GUT}^2$ as above.
Hence, to ensure a massless electroweak pair $h_u,\bar h_d$, we must impose the geometric mean relation
$M_{GUT}\sim\sqrt{M_S\,|v_R|}$.

In the more general situation where there are several Higgses
 (as is often the case in realistic constructions), (\ref{4H}) generalizes to
\bea
{\cal W}_H&\supset& \sum_{i,j}\bar H_j\,\Phi_4\,H_i+M_{H_{ij}}\,\bar H_i\,H_j\nn\\
&+&\sum_{i,j}\bar H_i'\,\Phi_4\,H_j'+M_{H_{ij}'}\,\bar H_i'\,H_j' \cdot
\label{4HB}
\eea
Now, the mass terms in (\ref{4HA}) become mass matrices  and one has to seek conditions for the existence of zero eigenvalues
in the ${\bar L}^c_{H_i},\,L_{H_i}^c$ mass matrix.

\section{ Pati Salam models at Gepner points}\label{PSGepners}

In the context of orientifolds constructed from Gepner models
there has been an extensive search for all possible embeddings of the SM gauge theory in D-brane
configurations~\cite{Anastasopoulos:2006da, Anastasopoulos:2009mr}\footnote{For some initial
studies of these constructions see~\cite{Angelantonj:1996mw}. }.
Among them, there are several cases where the SM is consistently embedded in a unified gauge group.
The predicted spectrum in these models includes all SM matter representations,
and the gauge couplings naturally unify  at scales $M_{GUT}\sim 10^{16}$GeV. Successful
candidate groups include $SU(5)$, flipped $SU(5)$, Pati-Salam gauge symmetry and
trinification models. Studies to identify semi-realistic unified models based
in $SU(5)$ symmetry had  appeared as well ~\cite{Anastasopoulos:2009mr}.
 In what follows, we elaborate on two characteristic examples with the PS gauge group
  chosen from  the pool of models derived in~\cite{Anastasopoulos:2006da, Anastasopoulos:2009mr}
 which have been found  in the context of Gepner constructions. Guided by our previous phenomenological
analysis, we pick up characteristic cases with massless spectra fulfilling most of
the aforementioned  requirements. First, we will deal  with a model possessing extra
states. This does not necessarily mean that the model is ruled out, however a number
of refinements are necessary in order to overcome some of the problems discussed previously. As
a second example, we  choose a model with a minimal spectrum which inherits several
of the nice features discussed above.

For both models we will present
a phenomenological analysis of the superpotential, the  fermion mass spectrum as well as
 related phenomenological issues. Doing so, we will need to make several mild assumptions about
 the parameter space.  We should not forget however, the rather subtle issue of  moduli stabilization
 \footnote{See~\cite{Blumenhagen:2006ci} and references therein.}, that might affect the freedom of choice for parameters used in this analysis. We will tacitly assume
  that  a stabilization mechanism does exist, so we will mainly concentrate on the general forms and  values
  for parameters in order to get semi-realistic realizations. We further note that
  there exists no explicit computation of Yukawas for orientifold models at Gepner points.
  This way, our superpotential is constrained  by gauge invariance only, without taking into account  conformal invariance restrictions that might shorten its length. This is not necessarily a shortcoming since several problematic gauge invariant couplings would probably be eliminated. 

\subsection{First Example}

Here, we present a vacuum with a Pati-Salam-like massless spectrum that consists of
the minimal configuration of three brane-stacks.
In this model, the internal sector consists of a tensor product of five copies of ${\cal N}=2$ superconformal minimal models with levels $k_i=3$ for $i=1,\dots , 5$. It contains a single orientifold plane.
A stack of 4 almost coincident branes
gives rise to the $U(4)$ symmetry, while  two stacks of 2 branes account for the two $U(2)$
gauge factors. The massless spectrum found together with the particle assignment
is as follows~\cite{Anastasopoulos:2006da}:
%
%
\bea
\begin{array}{llrcl}\nonumber
& Gauge~Group &~~~~~~~~~~~~ Chirality  & ~~~~~~~~ & Spectrum\\
& U(4)\times U(2)\times U(2) \\
3 \times & ( V ,0 , \bar V) & -1 & \to & \bar F_R  + H + \bar  H    \\
2 \times & ( V ,0 ,V ) & -2 & \to & 2  \bar F_R '     \\
1 \times & ( 0 ,0 ,S ) & 1 & \to &  \D_R   \\
5 \times & ( 0 ,A ,0 ) & 1 & \to & 3 \n +  2  \bar \n   \\
5 \times & ( V , \bar V,0 ) & 1 & \to & 3  F_L + 2 \bar F_L   \\
6 \times & ( V ,V ,0 ) & 2 & \to & 4 F_L' + 2 \bar F_L '   \\
3 \times & ( 0 ,V ,V ) & -1 & \to & h  + 2  \bar h \\
4 \times & ( 0 ,S ,0 ) & 0 & \to & 2 \Delta_L  + 2 \bar \Delta_L  \\
4 \times & ( S ,0 ,0 ) & 0 & \to & 2 S_{10}  + 2 \bar S_{10}  \\
3 \times & ( Adj,0 ,0 ) & 0 & \to & 0 \\
5 \times & ( 0 ,Adj,0 ) & 0 & \to & 0 \\
1 \times & ( 0 ,0 ,Adj) & 0 & \to & 0 \\
2 \times & ( 0 ,V , \bar V) & 0 & \to &  h'  + \bar h'
\end{array}
\eea
In the above,  the symbol $Adj$ denotes the adjoint, and $A,~S,~V$  the antisymmetric, the symmetric
and the fundamental representation of the relevant group respectively.  The ``bar'' refers to the conjugate representations.
 The charge of the above representations under the corresponding abelian factor of each stack is
 $2,~2,~1$ and $-2,~-2,~-1$ for the conjugate representations (see
also Table 1 for a detailed presentation of all quantum numbers).
Comparing the  Gepner vacuum above with Table \ref{U422}  we find a coincidence for
 the following choice of multiplicities: $n=2$, $m=2$, $\bar n=1$, $\bar m=0$.
In addition, we find the $\Delta_L$,  $\bar \Delta_L$, $S_{10}$,  $\bar S_{10}$ representations,
while the sextet $D_6$ field in this particular Gepner model is absent. This is a welcome
 fact since one can now avoid various awkward mixings with
SM fields, as discussed in the previous sections.

The number  multiplying the representation denotes the total number of states that appear in the spectrum.
In order to avoid too much clutter,  we present the spectrum without distinguishing between the fields and their conjugates.
Instead, in the column  ``chirality" of the above Table we designate the number of chiral fields.
For example,  there are 5 fields like $( V , \bar V,0 )$ in total but 4 of them appear like $( V , \bar V,0 )$, and
one with the opposite chirality $(  \bar V ,V,0 )$.

Next, we discuss the superpotential terms. Since we have extra matter pairs $\bar F_LF_L$, $\bar F_RF_R$,
it is natural to expect that in a viable model, they get a large mass.
This can be achieved by  an (effective) singlet non zero vev $\langle\phi_0\rangle $ or NR-terms with mass parameter
of the form $\langle\bar HH\rangle/M_S$.
Omitting  the multiplicity  indices for the fields $F_{L,i}$, $\bar F_{L,i}$, $F'_{L,i}$, $\bar F'_{L,i}$, $F_{R,i}$, $S_{10,i}$, $\bar S_{10,i}$, $\n_{i}$, $\bar \n_{i}$, $h_{i}$, $\Delta_{L,i}$, $\bar \Delta_{L,i}$, the bilinear
terms are
\bea
{\cal W}_2&=&
M\bar F_L F_L + M'\bar F_L' F_L' +\mu_1 \bar F_R
   H + \mu_2\bar H H \nn\\
   &&
   + \mu_S\bar S_{10} S_{10} +\mu_{\nu} \bar \n \n +\mu_s \bar \Delta_L
   \Delta_L + \mu'\bar h' h' + \mu\bar h h \cdot \label{w2}
\eea
Clearly, the natural scale of the mass parameters in (\ref{w2}) is of the order of $M_{GUT}$. Since
several of these couplings play a  decisive r\^ole for the viability of the model,  we discuss these
terms separately.

 The mass terms for the left-handed representations
$M \bar F_L F_L +M' \bar F_L' F_L' $ generate masses for two pairs of $F_L,\bar F_L$ fields
and another two of   $\bar F_L', F_L' $. Counting the total number of these  states and their
conjugates, we conclude that two linear combinations of $F_L'$ and another one from
$F_L$ remain massless. Thus, there are in total three massless ($4,2,1$)'s
  which suffice to accommodate the left-handed fields of the Standard Model.

Next, we discuss the couplings $\mu_1 \bar F_R H + \mu_2\bar H H $. We observe
that the appearance of the term $\bar F_R H $ indicates that
this model suffers from a shortcoming since
it fails to discriminate between
the Higgs and the RH $\bar F_R$ states. Indeed, one $\bar F_R$ necessarily shares the same $U(1)$
charges  with the $SU(4)$ breaking Higgs $\bar H$. In this case,  we may redefine the
fields $\bar F_R, \bar H$, so that the Higgs field is represented by
$\bar H''=\bar H\cos\theta   + \bar F_R\sin\theta $ with $\tan\theta =\frac{\mu_1}{\mu_2}$. Then,
the orthogonal linear combination to $\bar H'' $ accommodates
one right-handed fermion generation, thus
\bea
\bar H''&=&~~\bar H\cos\theta   + \bar F_R\sin\theta\label{hp}\\
\bar F_R''&=&-\bar H\sin\theta   + \bar F_R\cos\theta \cdot \label{fp}
\eea
Under this redefinition of fields, the two terms  `merge' into a Higgs coupling $M_H \bar H''H$ with
$M_H=\sqrt{\mu_1^2+\mu_2^2}$ and the analysis can be carried out just as discussed in the previous sections.

 Analogous mass terms appear in (\ref{w2}) for other  states.  Among those remaining terms, a separate discussion should
 be devoted to the bidoublet fields because of their crucial r\^ole in the electroweak symmetry
 breaking. The two terms $\mu'\bar h' h' + \mu\bar h h$ render two pairs of bidoublets massive, thus,
consulting the table representing the spectrum of the model we conclude that there always remains  one bidoublet $\bar h$
in the massless spectrum,  up to this order at least.
Actually, the problem of the Higgs mass is even more complicated
since one is confronted with affluent doublets and Yukawa couplings between them in these constructions.
Consequently, in order to determine the massless Higgs spectrum
a more detailed analysis should be carried out, including higher order superpotential terms.
 This is usually feasible (see, for example, \cite{Ellis:1999ce})
by  appropriately tuning the unknown parameters (i.e., Yukawa couplings and various
vev scales). However, such an analysis goes beyond the scope of the present work. Alternatively,
one may implement the splitting scenario discussed in the previous sections to ensure the existence of a massless $SU(2)_L$ Higgs pair
$h_u',\bar h_d'$.

 We now proceed  to the trilinear couplings. Among them, the most important  are those which provide
 masses to ordinary fermions. The terms supplying the families with Dirac masses are
\bea
{\cal W}_{3}&\supset &(\lambda_{1F}  \bar F_{R}+\lambda_{1H}\bar H ) F_L h' +
                       (\lambda_{2F} \bar F_{R}+\lambda_{2H}\bar H ) F_{L,i}'  \bar h  +
\lambda_3 F_L  \bar F_{R,i}'h + \lambda_4 \bar F_{R,i}' F_{L,i}'\bar h'\nn
\eea
where $\lambda_i$ stand for the Yukawa couplings. Under the redefinitions (\ref{hp},\ref{fp}) the above
terms imply the following fermion generation Yukawa couplings
\bea
{\cal W}_{3}&\supset &\lambda_1' \bar F''_RF_Lh'  + \lambda_2' \bar F''_{R}F_{L,i}'  \bar h  +
\lambda_4 F_L  \bar F_{R,i}'h + \lambda_5 \bar F_{R,i}' F_{L,i}'\bar h'\label{modY}
\eea
with $\lambda_j'=\lambda_{jF}\cos\theta-\lambda_{jH}\sin\theta$ and $j=1,2$.

Before analyzing  the resulting mass terms (\ref{modY}), we point out that the above field rotation generates
also the couplings
$\l''_1 \bar H''F_Lh'+\l''_2\bar H''F'_L\bar h$, with $\l''_j=\lambda_{jF}\sin\theta+\lambda_{jH}\cos\theta$ where $j=1,2$.
Clearly, since the $\bar H''$ field acquires a GUT order vev, a corresponding mass
term is generated for each of the lepton doublets coupled  to
the appropriate doublets $h'_u\in h'$ and $\bar h_u\in \bar h$.  A definite conclusion on the masslessness of
the lepton doublet fields would require the investigation of the complete doublet mass matrix and the
determination of the mass eigenstates. A simpler approach to this problem would be to assume that there is enough
freedom in the parameter space, so that one can impose the condition $\l''_j=0$, (or
 in terms of the mass parameters $\mu_1\l_{jF}+\mu_2\l_{jH}=0$) and the
problematic couplings disappear from the superpotential.
Additional redefinitions may  be applied to the  left-handed fields since these fields  participate also in
couplings which involve pairs $\bar F_LF_L$ and so on. Nevertheless, such additional effects will not modify
 our general discussion and for our present purposes such redefinitions will not be considered.

 To estimate the individual contributions coming from each one of the above
terms in the fermion mass matrix, we first discuss the bidoublet Higgs spectrum. We have previously seen  that
bilinear mass terms  leave in general only one massless bidoublet state, namely one $\bar h=(1,2,2)$. The masslessness
of the Higgs field(s), can only be ensured  if an inspection of the non-renormalizable  terms up to a sufficient order
is consistently carried out ~\cite{Ellis:1999ce}.  When constructing the entire bidoublet mass matrix,
we expect that any other contribution is  hierarchically smaller compared to the bilinear mass terms in (\ref{w2}). Therefore,
 it is natural to expect that the main component of the bilinear Higgs combination $\alpha_i\,h_i+\beta_j\,\bar h_j$
 which will eventually remain massless will arise from $\bar h(1,2,2)$.  We infer that
 the large entries in the fermion mass matrices come from the terms  $\lambda_2' \bar F''_{R}F_{L,i}'  \bar h$
 in (\ref{modY}) and therefore, $\bar F''_{R},F_{L,i}'$ are suitable for accommodating the heavier generations.
 Motivated by the above, we define the parameters $\epsilon_1\sim\frac{\lambda_1' h'}{\lambda_2'\bar h}\ll 1$, etc.,
thus it is expected that the general structure of a typical mass matrix in this model obtains the form
 \bea
M_{u,d,l} &\sim&~\left(
\begin{array}{lll}
\e_{13} ~ &\e_{23} ~ &\e_1  \\
\e_{14} ~ &\e_{24} ~ &\lesssim 1  \\
\e_{34} &\e_{44} &1\\
\end{array}\right) \cdot
\eea
Since $\e_{ij}\ll 1$, this texture roughly predicts the anticipated hierarchical fermion mass pattern.

We finally come to the neutrino sector. To realize the see-saw mechanism and bring the
masses down to experimentally acceptable limits, we need to search for heavy Majorana contributions
to the right-handed partners. The  gauge  invariant couplings we find at the tree-level as well as
at the fourth order are
\bea
{\cal W}_{NR}&\supset&\kappa_0\,H\bar F'_{R_j}\Delta_R+\left(\k_1 \bar F_R^2+\k_2\bar F_R\bar H+\kappa_3\bar H^2\right)\,H^2\nn\\
 &\ra&\kappa_0\,H\bar F'_{R_j}\Delta_R+\kappa\,(\bar F''_R)^2\,H^2+\cdots
\eea
The first term which couples the ordinary RH-neutrinos to the neutral component of $\Delta_R$ was already discussed
in the previous sections. In the second line, the couplings were written in terms of the redefined field $\bar F''_R$
and dots stand for terms $(H\,\bar H'')^2$ etc.
In the basis $\bar F_{R_j}',\bar F_R'',\Delta_R$, the extended heavy neutrino mass matrix becomes (suppressing for
convenience the index $j=1,2$)
\bea
M_{\nu^c}&=&\left(
\begin{array}{ccc}
 0 & 0 & M_{GUT} \\
 0 & \frac{M_{GUT}^2}{M_S} & 0 \\
 M_{GUT} & 0 & 0
\end{array}
\right) \cdot
\eea
Here, $M_{GUT}\sim \langle H\rangle$ and $M_S$ is an effective (string) scale suppressing the fourth order Yukawa couplings.
Contributions from higher order corrections to the mass matrix are of course expected, but the essential result
does not change. The form of the heavy neutrino mass matrix is appropriate for suppressing the left-handed neutrino
masses down to the desired level, through the  see-saw mechanism.

\subsection{Second example}

One of the semi-realistic Gepner constructions presented in~\cite{Anastasopoulos:2006da} is based on an extended  PS gauge symmetry by two additional $U(2)$ factors, thus the corresponding  gauge symmetry is
$$ \left[U(4)\times U(2)\times U(2)\right]_{\rm obs.} \times \left[U(2) \times U(2)\right]_{\rm hid.}$$

In this model, the internal sector consists of a tensor product of five copies of ${\cal N}=2$ superconformal minimal models with levels $k_i={1,1,2,14,46}$, and
it contains a single orientifold plane.

The massless fields read as follows:
%
%
\bea
\begin{array}{llrcl}\nonumber
& Gauge~Group & Chirality  & ~~~~~~~~ & Spectrum\\
&      U(4)\times U(2)\times U(2) \times U(2) \times U(2)\\
      3 \times & ( V ,V ,0 ;0 ,0 ) &   3     & \to &  F_{L\,i},\, i=1,2,3\\
      4 \times & ( V ,0 , \bar V;0 ,0 ) &   0     & \to &  H_a +  \bar H_{a}, \,a=1,2\\
      3 \times & ( V ,0 ,V ;0 ,0 ) &   -3    & \to &  \bar F_{R\,i}'\\
      7 \times & ( 0 ,0 ,S ;0 ,0 ) &   3     & \to & 5 \Delta_R+ 2 \bar \Delta_R\\
      8 \times & ( 0 ,S ,0 ;0 ,0 ) &   0     & \to & 4 \D_L + 4 \bar \D_L \\
      3 \times & ( 0 ,A ,0 ;0 ,0 ) &   3     & \to & 3 \n\\
      3 \times & ( 0 ,V ,V ;0 ,0 ) &   -3    & \to & 3 \bar  h\\
      4 \times & ( 0 ,V , \bar V;0 ,0 ) &   0     & \to & 2h' + 2 \bar h'\\
      1 \times & ( Adj,0 ,0 ;0 ,0 ) &   0\\
      2 \times & ( 0 ,Adj,0 ;0 ,0 ) &   0\\
      3 \times & ( 0 ,0 ,Adj;0 ,0 ) &   0\\
      2 \times & ( 0 ,0 ,0 ;0 ,Adj) &   0\\
      6 \times & ( 0 ,V ,0 ;0 ,V ) &   0\\
      2 \times & ( 0 ,0 ,V ; \bar V,0 ) &   0\\
      \end{array}\eea
The hypercharge embedding is
$Y={1\over 6}A_3+{1\over 2}A_1+{1\over 2}A_1'$
while all the remaining abelian factors are massive due to anomalies.

This is a rather interesting case from the point of view that there are no extra vector-like pairs
sitting in the same representations with  ordinary families. Therefore, `creation' of tree-level
terms $M_{GUT}\bar F_iF_i$ which would also eliminate the Higgs can be avoided and the
 analysis can be simpler.

The  trilinear terms are:
\bea
 {\cal W}_3 &=&\bar H F_L h' + \bar F_R' \D_R H + \bar h \bar h' \nu +  \bar h \bar h' \D_L +  \bar h \D_R h' \cdot
  \eea
 There are no fermion mass terms at this level, so these are expected to appear from higher order terms.  In addition,
in order to avoid severe problems due to the presence of the $\bar H F_L h'$ term, we demand that $\langle\,h'\rangle=0$.
 The $SU(4)$ breaking Higgs acquires a non-vanishing vev $\langle \bar H \rangle\ne 0$, combining
 the lepton doublet in $F_L$ with the appropriate  $SU(2)_L$ doublet in $h'\ra h_u'+h_d'$ to a heavy
 mass term $\sim M_{GUT}\ell h_u'$. In this case, the $SU(2)$ lepton doublet of the particular family should be accommodated in
 the remaining massless part of $h'$ bidoublet, i.e., $h_d'\equiv \ell$.
 Looking for fermion mass terms at higher orders,  we get from fourth order NR-contributions
 \bea
 {\cal W}_4 &\supset& \bar F_R'  F_L (\bar h' \nu +  \D_R  h') \cdot \label{fmt}
\eea
Since  $h'$ has a vanishing vev,  at least one of the two bidoublets $\bar h'$ has to acquire
a non-vanishing vev, $\langle\bar h'\rangle\ne 0$.
Furthermore, to generate mass terms at the fourth order, we must turn on a non-zero vev for the neutral singlet
$\nu$.
 We may now write  the relevant term  of (\ref{fmt}) giving masses to the fermions
  \bea
{\cal W}_Y&\sim& \lambda_{ij}\bar F_{Ri}'  F_{Lj} \langle \hat h \nu \rangle\label{fmt1}
\eea
where the indices $i,j$ take all the values $i,j=1,2,3$.
In addition, we can search for heavy RH-neutrino contributions, which realize the see-saw mechanism. Already
at the fourth order we can find the term
\bea
 H \bar \D_R \bar F_R \D_R  &\rightarrow&\langle H\bar \D_R\rangle\,\nu^c\delta^0_R \cdot
\eea
Similar contributions are naturally expected to occur in higher order NR-terms. These contributions will lead to
a heavy RH neutrino mass matrix coupled to other neutral states and can prove sufficient
to suppress the LH neutrino masses and reconcile the data.

\subsubsection{A variation of model 2}
When analyzing model 2 in the last section, we noticed that all Yukawa couplings are realized
at the fourth order, since no-tree level mass terms can exist. This situation may look uncomfortable
in the sense that the top-quark Yukawa coupling is also derived from a fourth order superpotential
term. Although this fact cannot necessarily prevent the model from reconciling the data, we would like to discuss
in brief an alternative interpretation of the spectrum which would result to a tree-level term.
To this end, we rename the fields of the previous case as follows:
%
%
\bea
\begin{array}{llrcl}\nonumber
& Gauge~Group & Chirality  & ~~~~~~~~ & Spectrum\\
&      U(4)\times U(2)\times U(2) \times U(2) \times U(2)\\
      3 \times & ( V ,V ,0 ;0 ,0 ) &   3     & \to &  F_{L\,i},\, i=1,2,3\\
      4 \times & ( V ,0 , \bar V;0 ,0 ) &   0     & \to &  H_a +  \bar F_{R\,a}, \,a=2,3\\
      3 \times & ( V ,0 ,V ;0 ,0 ) &   -3    & \to &  \bar F_R'+ \bar H'_a\\
      \end{array}\eea
where only the relevant spectrum is shown, since the remaining fields do not change.
The  trilinear terms are:
\bea
 {\cal W}_3 &=&\bar F_R F_L h' + (\bar F_R' + 2 \bar H') \D_R H + \bar h \bar h' \nu +  \bar h \bar h' \D_L +  \bar h \D_R h'
  \eea
%
%
%
%
and imply an interesting structure for the mass matrices
\bea
m_{u,d,l,\nu}\sim\left(
\begin{array}{lll}
0 & 0&0 \\
\lambda_{21} &\lambda_{22} & \lambda_{23} \\
 \lambda_{31} &\lambda_{32} & \lambda_{33}
\end{array}
\right)\langle h'\rangle \cdot
\label{ca3}
\eea
Such  textures seem to be rather generic in several D-brane constructions.
If these were to be the only contributions, the model would be ruled out for
predicting zero  masses  for the light generation. However, the zeros in the
above textures could be filled  with either non-renormalizable
terms or instanton contributions as discussed in section (\ref{Instantsection}).
Assuming that such contributions exist, the mass matrix takes the form
\bea
m_{u,d,l,\nu}\sim\left(
\begin{array}{lll}
h_{11} & h_{12}&h_{13} \\
\lambda_{21} &\lambda_{22} & \lambda_{23} \\
 \lambda_{31} &\lambda_{32} & \lambda_{33}
\end{array}
\right)\langle h'\rangle
\label{ca3-1}
\eea
where $h_{1i}, i=1,2,3$ are the relevant Yukawa couplings, and it is expected that $h_{1i}\ll \lambda_{jk}$.
These suppressed contributions are  in accordance with the family mass hierarchy. Nevertheless, the predicted
structure in this model is rather peculiar. Indeed, we have seen that the  last two lines of the matrix
receive contributions at the tree-level, and thus it is naturally   expected that the scales of their entries
are comparable, i.e., $\lambda_{2i}\sim {\cal O}(\lambda_{3j})$ at least for some values of the indices
$i,j$. On the other hand, the question arises whether such a non-symmetric
mass texture can still  reconcile the experimentally known pattern of fermion generations. Although at
present we do not know how to calculate the exact values of Yukawas for a given model, we stress that this
rather peculiar structure is at least compatible with the observed  fermion mass hierarchy. Since this
is decisive for the viability of the model, we will devote a separate discussion on the analysis
of the above texture.
To this end,  we define the vectors $\vec\xi_j$, $j=1,2,3$,
\bea
\vec{\xi}_1=(h_{11},h_{12},h_{12}),\;\;
\vec{\xi}_k=(\lambda_{k1},\lambda_{k2},\lambda_{k3}), k=2,3
\eea
so that the generic form of the above matrices is written in the vector like form
\bea
m_{\vec\xi}= ( \vec{\xi}_1 ,\vec{\xi}_2 , \vec{\xi}_3)^T \cdot
\eea
We argue that this
vector-like presentation of the matrix is the most appropriate for investigating the
viability of  textures like (\ref{ca3}). Indeed, we have seen that the first line of
this  matrix as compared to the other two
 is characterized by a vastly different mass scale.
Instead, therefore, of seeking solutions for individual Yukawa couplings, it is adequate
to investigate the viability conditions for a structure with hierarchy
\bea
|\vec\xi_1|\;\;\ll \;\;|\vec\xi_2|\;\;\sim\;\; |\vec\xi_3| \cdot \label{vech}
\eea
To simplify the analysis, we can bring the matrix $m_{\vec\xi} $ into a lower triangular (Cholesky)
form $m_C$ and work out cases with real entries\cite{Leontaris:2009pi}.
The two matrices are connected by an orthogonal matrix $U$, i.e., $m_{\vec\xi}=m_C\cdot U$, where
$U=(\hat{e}_1,\hat{e}_2,\hat{e}_3)^T$ (with $\hat{e}_i$ representing three-vectors),
or,  analytically
\bea
m_{\vec\xi}\equiv\left(
\begin{array}{c}
\vec{\xi}_1  \\
\vec{\xi}_2 \\
\vec{\xi}_3
\end{array}
\right)&=&\left(
\begin{array}{ccc}
\vec\xi_1\cdot\hat e_1  &0&0\\
\vec\xi_2\cdot\hat e_1  &\vec\xi_2\cdot\hat e_2  &0 \\
\vec\xi_3\cdot\hat e_1  &\vec\xi_3\cdot\hat e_2  &\vec\xi_3\cdot\hat e_3
\end{array}
\right)\left(
\begin{array}{c}
\hat{e}_1  \\
\hat{e}_2 \\
\hat{e}_3
\end{array}
\right) \cdot \label{Chol}
\eea
We can easily observe that the transformation of the original matrix to its Cholesky form does not
affect the eigenvalues and eigenvectors , since
\bea
m_{\vec\xi}\cdot m_{\vec\xi}^T&=\;(m_C\cdot U)\cdot (U^T\cdot m_C^T)&\equiv \;m_Cm_C^T \cdot \label{class}
\eea
It can be shown~\cite{Leontaris:2009pi} that all the elements of $m_C$ can be expressed as simple
functions of the mass eigenstates and the diagonalizing matrix entries,  the latter being
related to the Cabbibo-Kobayashi-Maskawa (CKM)   mixing effects.
 Thus, using the triangular form of the matrix where everything can
be expressed in terms of physical quantities (masses and mixing), we may easily seek solutions
that satisfy the required inequality (\ref{vech}). Returning to our particular texture,
 we give an  illustrative example for the case  of
the quark sector where the quark masses and the CKM matrix are experimentally known
to a good precision. Let therefore, the down quark mass texture be~\cite{Leontaris:2009pi}
\bea
m_D&=&\left(
\begin{array}{ccc}
m_d & 0 & 0 \\
0 & \sqrt{m_s^2 \cos ^2\beta+m_b^2 \sin ^2\beta}& 0 \\
0 & \frac{(m_b^2-m_s^2) \sin (2\beta)}{2\sqrt{m_s^2 \cos ^2\beta+m_b^2 \sin ^2\beta}} & \frac{m_b
 m_s}{\sqrt{m_s^2 \cos ^2\beta+m_b^2 \sin ^2\beta}}
\end{array}
\right)
\eea
which can be checked to give the correct down quark masses for any value of the arbitrary angle $\beta$.
To keep the algebra  tractable, we have assumed a Cholesky texture with $m_{21}=m_{31}=0$,
however, as it can be seen from (\ref{class}), there is a whole class of mass matrices
 $m_C\cdot U$, where $U$ is any orthogonal
matrix, which have the same physical properties (mass eigenstates and mixing).
Clearly, $m_D$ was chosen in a way so that the condition $ |\vec\xi_1|\ll |\vec\xi_{2,3}|$ is satisfied,
while we can still adjust the value of
$\beta$ to obtain the naturalness  condition $ |\vec\xi_2|\sim |\vec\xi_{3}|$, since
 the components of both vectors are related to tree-level Yukawa couplings. Choosing, for example, $\beta=\frac{\pi}{4}$
while substituting the quark masses we obtain
\bea
m_D&=&
\left(
\begin{array}{lll}
0.005 & 0 & 0 \\
0 & 3.01 & 0 \\
0 & 3.001 & 0.177
\end{array}
\right)\,m_D^0 \cdot
\eea
This matrix clearly satisfies  the required conditions.
Using the well known CKM matrix we can  calculate the up-quark mass matrix which takes the form
\bea
m_U&=&
\left(
\begin{array}{lll}
0.29 & 0 & 0 \\
43.23 & 107.4 & 0 \\
48.6 & 116.1 & 0.02
\end{array}
\right)\,m_U^0
\eea
and exhibits  the same `vector' hierarchy (\ref{vech}) as the down quark mass matrix. This example suggests
 that,  in principle at least, it may be feasible to obtain a correct hierarchical mass spectrum and
CKM mixing from the predicted mass texture (\ref{ca3}).

These were just two characteristic examples picked up from a wider number of models derived in~\cite{Anastasopoulos:2006da}.
Several PS models with different spectra  (some also with additional gauge group factors
in the hidden sector of the theory)  are collected in the Appendix for reasons of completeness.

\section{Conclusions}\label{Concl}

In this work, we  performed a generic phenomenological analysis of the effective low energy models with Pati-Salam (PS)  gauge symmetry
derived in the context of D-brane vacua,  concentrating on the major issues.
We discussed the problem raised by the absence of  Yukawa couplings prevented by $U(1)$ symmetries, and suggested solutions to generate the
missing terms. We analyzed the implications of  the various exotic representations which can appear in the D-brane PS models
and presented viable scenarios to decouple them from the light spectrum. In these string vacua, the right-handed
fermions and the PS-breaking  Higgs fields are described
by the same kind of strings stretched between the $U(4)$ and $U(2)_R$ D-brane stacks.
This fact, typically leads to undesirable couplings,
which put the reliability of these models under question. We  showed that  this  problem can be bypassed by focusing on classes of models where
the right-handed fermions emerge from strings attached to $U(2)_R$ stack while the  PS  breaking Higgs fields are described by
strings attached to its mirror.  In this case,  the presence of a heavy Higgs mass term eliminates all the unwanted effects
from the spectrum, by means of a doublet-triplet splitting mechanism and  an alternative symmetry breaking pattern.

Furthermore, we analyzed the mass matrices that appear in these models and  argued on the importance of higher order non-renormalizable
terms and the stringy instantonic contributions that generate the missing Yukawa couplings  contributing to the fermion mass textures.
We also described how in certain cases the antisymmetric and symmetric representations can trigger the see-saw mechanism,
to generate the light neutrino masses.

In addition, we discussed the correlations between the intersecting D-brane spectra and those obtained from Gepner constructions
and analyzed the superpotential, the  mass textures and the low energy implications in two examples
derived in~\cite{Anastasopoulos:2006da}.

The first one is a  typical case  in a characteristic  class of Gepner models with minimal PS symmetry.
The spectrum is rather complicated  and contains several exotic states.
We have explored mechanisms to remove the exotics from the light spectrum and discussed ways to obtain a viable low energy effective model.
The second  model is characterized by an extended PS symmetry and additional $U(2)$ hidden gauge factors.
It  has exactly three families without extra vector like pairs, and is free from  exotic representations.
Two variants of this case were presented and found that
there are definite predictions for the fermion mass textures which, in principle, can be compatible
with the low energy fermion mass data. This is an encouraging fact which prompts for further future investigation.

\vspace*{1cm}

\begin{center}
{\bf Aknowledgements}
\end{center}

We would like to thank Ignatios Antoniadis, Elias Kiritsis, Bert Schellekens
and in particular Robert Richter, for  illuminating discussions, useful comments and reading the manuscript.

P.A. would like to thank University of Crete, LPT Universit\'e Paris XI Orsay and LP Ecole Normale Sup\'erieure de Lyon for hospitality during the last stage of this work.

{\it This work is partially supported by the European Research and Training
Network  `Unification in the LHC era' (PITN-GA-2009-237920).}

\newpage

\section*{\Large{Appendix}}

\bigskip\appendix

\section{Superpotentials}

Here we give for completeness the tree-level and  the fourth order superpotential  terms of
the models discussed in the text. Unless explicitly written, fourth order terms are assumed to be divided by the scale $M_S$.

\subsection{First example}
\bea
{\cal W}_3&=&
 \bar \Delta_L h h' + \bar \n h h' + \bar h \Delta_L
   \bar h' + \bar H F_L h' + \bar h \bar H
   F_L' + \bar F_L H \bar h' \nn\\
&&+ \bar F_L \bar \Delta_L F_L' + \bar F_L \bar \n F_L' + h H \bar F_L' + F_L
   \Delta_L \bar F_L' + \bar F_R F_L h' + \bar F_R \bar h
   F_L' \nn\\
&&+ F_L h \bar F_R' + \bar F_R' \bar h'
   F_L' + \bar S_{10} F_L F_L' + \bar F_L S_{10}
   \bar F_L' + \bar H S_{10} \bar F_R' \nn\\
&&+ \bar F_R S_{10}
   \bar F_R' + \bar h \D_R h' + \D_R H
   \bar F_R' + \bar h \n \bar h' + F_L \n \bar F_L'\nn\\
{\cal W}_4&=&
 \bar h^2 h^2 + \bar h \bar H h H + \bar H^2
   H^2 + \bar h h \bar h' h' + \bar H H \bar h'
   h' + \bar h'^2 h'^2 + \bar h \bar \Delta_L
   h \Delta_L \nn\\
&&+ \bar H \bar \Delta_L H \Delta_L + \bar \Delta_L \Delta_L
   \bar h' h' + \bar \Delta_L^2 \Delta_L^2 
   + \bar h \bar H F_L \Delta_L + \bar H
   \bar \Delta_L F_L' h' + \bar H \bar \n F_L' h' \nn\\
&&+ \bar F_L \bar \Delta_L h H + \bar F_L \bar \n h H + \bar F_L
   \bar h F_L h + \bar F_L \bar H F_L H + \bar F_L
   F_L \bar h' h' + \bar F_L \bar \Delta_L F_L
   \Delta_L \nn\\
&&+ \bar F_L \bar h \bar h' F_L' + \bar F_L^2
   F_L^2 + H \Delta_L \bar F_L' \bar h' + F_L h
   \bar F_L' h' + \bar h h \bar F_L' F_L' + \bar H H
   \bar F_L' F_L' + \bar F_L' \bar h' F_L'
   h' \nn\\
&&+ \bar \Delta_L \Delta_L \bar F_L' F_L' + \bar F_L F_L
   \bar F_L' F_L' + \bar F_L'^2
   F_L'^2 + \bar F_R \bar h h H + \bar F_R
   \bar H H^2 + \bar F_R H \bar h' h' \nn\\
&&+ \bar F_R \bar \Delta_L H
   \Delta_L + \bar F_R \bar h F_L \Delta_L + \bar F_R
   \bar \Delta_L F_L' h' + \bar F_R \bar \n F_L'
   h' + \bar F_L \bar F_R F_L H + \bar F_R H \bar F_L'
   F_L' \nn\\
&&+ \bar F_R^2 H^2 + h H \bar F_R' \bar h' + F_L
   \Delta_L \bar F_R' \bar h' + \bar \Delta_L h \bar F_R'
   F_L' + \bar \n h \bar F_R' F_L' + \bar H F_L
   \bar F_R' F_L' + \bar F_R F_L \bar F_R'
   F_L' \nn\\
&&+ \bar S_{10} F_L h H + \bar S_{10} F_L^2
   \Delta_L + \bar S_{10} H \bar h' F_L' + \bar S_{10}
   \bar \Delta_L F_L'^2 + \bar \n \bar S_{10}
   F_L'^2 + \bar F_L \bar h \bar H S_{10} \nn\\
&&+ \bar F_L^2 \bar \Delta_L S_{10} + \bar F_L^2
   \bar \n S_{10} + \bar H S_{10} \bar F_L' h' + S_{10}
   \Delta_L \bar F_L'^2 + \bar F_L \bar F_R
   \bar h S_{10} + \bar F_R S_{10} \bar F_L'
   h' \nn\\
&&+ \bar F_L S_{10} \bar F_R' \bar h' + h S_{10}
   \bar F_L' \bar F_R' + \bar h \bar S_{10} h
   S_{10} + \bar H \bar S_{10} H S_{10} + \bar S_{10} S_{10}
   \bar h' h' + \bar S_{10} \bar \Delta_L S_{10}
   \Delta_L \nn\\
&&+ \bar F_L \bar S_{10} F_L S_{10} + \bar S_{10}
   S_{10} \bar F_L' F_L' + \bar F_R \bar S_{10} H
   S_{10} + \bar S_{10}^2 S_{10}^2 + \bar \Delta_L \D_R h'^2 + \bar \n \D_R h'^2 \nn\\
&&+ \bar h^2
   \D_R \Delta_L + \bar F_L \bar h \D_R H + \D_R H
   \bar F_L' h' + \D_R F_L \bar F_R' h' + \bar h
   \D_R \bar F_R' F_L' + \bar S_{10} \D_R H^2 \nn\\
&&+ \D_R
   S_{10} \bar F_R'^2 + \bar h \bar \n h
   \n + \bar H \bar \n H \n + \bar \n \n \bar h' h' + \bar \n
   \bar \Delta_L \n \Delta_L + \bar h \bar H F_L \n + \bar F_L
   \bar \n F_L \n \nn\\
&&+ H \n \bar F_L' \bar h' + \bar \n \n
   \bar F_L' F_L' + \bar F_R \bar \n H \n + \bar F_R
   \bar h F_L \n + F_L \n \bar F_R'
   \bar h' \nn\\
&&+ \bar S_{10} F_L^2 \n + \n S_{10}
   \bar F_L'^2 + \bar \n \bar S_{10} \n
   S_{10} + \bar h^2 \D_R \n 
   + \bar \n^2 \n^2
\label{potential}\eea
where ${\cal W}_3$ denotes the cubic and ${\cal W}_4$ the nonrenormalizable terms.

Notice  that, several terms which are present in the Pati-Salam model are now absent at tree level due
to conservation under the extended $SU(4)\times SU(2)\times SU(2)\times U(1)\times U(1)\times U(1)$ gauge group. For example, the term $HHD_6$  does not
preserve U(1) gauge invariance: $(1,0,-1)+(1,0,-1)+(2,0,0)=(4,0,-2)$ and consequently, is not
present in (\ref{potential}). Such terms can be present due to instanton effects fig.~\ref{flr}.

\subsubsection{Flatness}

In model 1 we  made a choice of vevs which is consistent with the F- and D-flatness
of the superpotential. For a superpotential ${\cal W}(\Phi_j)$, where $\Phi_j$ stands for
the various superfields, the F-flatness conditions read
\bea
\frac{\partial {\cal W}}{\partial  \Phi_j}&=&0
\eea
The fields with non-zero vevs should satisfy also the D-flatness conditions. For the non-anomalous
$U(1)$'s
\bea
\sum_{j}Q_j |\Phi_j|^2&=&0
\eea
and for the anomalous ones
\bea
\sum_{j}Q_j^A |\Phi_j|^2&=&-\frac{\rm {Tr}Q_j^A}{192\pi^2}\,g_s^2M_{S}^2
\eea
where $Q_j,Q_j^A$ the corresponding $U(1)$ charges of the field $\Phi_j$.

Since in this case we have extra vector multiplets $\bar F_LF_L$, $\bar S_{10}S_{10}$ etc, we may assume that an
appropriate  singlet or other (i.e., $U(4)$-adjoint)  vev generates mass terms of the form\footnote{Such mass terms may also
be generated non-perturbatively by stringy instanton effects,  $\mu_i\sim e^{-S_{E_i}}\,M_S$ where
$S_E$ is the instanton action and  $M_S$ is the string scale.
A reasonable value for the parameters would be $\mu_i\sim 10^{-1}M_{GUT}\sim 10^{-2}M_S$ which implies
 a rather plausible value for the instanton suppression factor  $e^{-S_{E_i}}\sim 10^{-2}$. }
\bea
\mu_H \bar H\,H+\mu_R\bar F_RH+\mu_{\nu}\bar\nu\nu+\mu_S\bar S_{10}S_{10}+\cdots
\eea
We first note that there are  two vastly different mass scales in the model,  $M_{GUT}\gg m_W$.
The high GUT scale is determined by the vacuum expectation values of the  $SU(4)$ breaking Higgses
$\langle H\rangle\sim M_{GUT}$ etc, whilst the weak scale is related to the vevs of the bidoublets
$\langle h,\bar h\rangle\sim m_W$. Therefore, to check  the consistency of the  choice of $SU(4)$ vevs,
we may omit small ${\cal O}(m_W)$ contributions to flatness.

We may assume $\langle \bar H\rangle \gg \langle h\rangle$
and  $\langle F_L\rangle=\langle \bar F_L\rangle =0$. We also  put $\langle F_R\rangle=0$
for all three representations accommodating the right-handed fermions
and the $F$-flatness conditions simplify to
\bea
\frac{\partial {\cal W}}{\partial  H}&\approx&\mu_H\bar H+
2\bar H^2  H + \bar H \bar S_{10} S_{10}  + \bar H \bar \D_L \D_L
+2\Delta_R\bar S_{10} H+\bar H\bar\nu\nu
\\
\frac{\partial {\cal W}}{\partial \bar H}&\approx&\mu_H H+ 2\bar H  H^2 + \bar H \bar \D_L \D_L
+  H\bar S_{10} S_{10}+H\bar\nu\nu
  \eea
Analogously, taking derivatives of the remaining fields we have the additional terms
\bea
\frac{\partial {\cal W}}{\partial  \bar F_R}&\approx&\mu_RH+
  \bar H H^2+ \bar \Delta_L H   \Delta_L +\bar \n H \n  + \bar S_{10} H  S_{10}
\\
\frac{\partial {\cal W}}{\partial   S_{10}}&\approx&\mu_{S}\bar S_{10}+\bar H\,H\bar S_{10}
+\bar S_{10}\bar \Delta_L\,\Delta_L+2\bar S_{10}^2\, S_{10}+\bar S_{10}\bar\nu\nu\\
\frac{\partial {\cal W}}{\partial  \bar S_{10}}&\approx&\mu_{S}S_{10}+\bar H\,H\, S_{10}
+ S_{10}\bar \Delta_L\,\Delta_L+2\bar S_{10}\, S_{10}^2+S_{10}\bar\nu\nu+\Delta_R\,H^2\label{FS10}\\
\frac{\partial {\cal W}}{\partial   \nu}&\approx&\mu_{\nu}  \bar \nu+\bar H\,H\bar\nu+S_{10}\bar S_{10}\bar\nu
+2\nu\bar\nu^2\\
\frac{\partial {\cal W}}{\partial  \bar \nu}&\approx&\mu_{\nu}\bar\nu+\bar H\,H\,\nu+S_{10}\bar S_{10}\,\nu
+2\nu^2 \bar\nu\\
\frac{\partial {\cal W}}{\partial  \Delta_L}&\approx&\mu_{s}\bar \Delta_L+2\Delta_L\bar \Delta_L^2+\bar \Delta_LS_{10}\bar S_{10}+\bar \Delta_L H \bar H
\\
\frac{\partial {\cal W}}{\partial \bar  \Delta_L}&\approx&\mu_{s}\Delta_L+2\Delta_L^2\bar \Delta_L+ \Delta_LS_{10}\bar S_{10}+ \Delta_LH\bar H
\\
\frac{\partial {\cal W}}{\partial \Delta_R}&\approx&\bar S_{10}H^2
\eea
In the above, we have omitted a common denominator $M_S$, dividing all fourth order terms.
Since $S_{10}$ fields carry charge and color, we must put   $\langle S_{10}\rangle=\langle\bar S_{10}\rangle=0$, while any non-zero vev for the left-handed triplets is negligible, thus
we also put $\langle \Delta_L\rangle=\langle\bar \Delta_L\rangle\approx 0$. Furthermore,
charge conservation implies that in (\ref{FS10}) $\langle \Delta_RH^2\rangle =0$, even
if both fields $\Delta_R$ and $H$ acquire non-zero vevs.
Plugging into the F-flatness conditions, while restoring units $M_S$ for the
 NR-terms, we arrive at the simple system of algebraic equations
\bea
2\langle \bar HH\rangle+\langle \bar\nu\nu\rangle+\mu_H\,M_S&\approx&0\\
\langle \bar HH\rangle+2\langle \bar\nu\nu\rangle+\mu_{\nu}\,M_S&\approx&0\\
\langle \bar HH\rangle+\langle \bar\nu\nu\rangle+\mu_R\,M_S&\approx&0
\eea
Thus, the F-flatness conditions are consistent with the choice of vevs
\bea
\frac{\langle H\bar H\rangle}{M_S}\sim\frac{\mu_{\nu}-2\mu_{H}}{3}&,&
\frac{\langle \bar\nu\nu\rangle}{M_S}\sim \frac{\mu_{H}-2\mu_{\nu}}{3}
\eea
while $\langle\Delta_R\rangle$ is not constrained by F-flatness. On the contrary,
 the condition $\mu_{H}+\mu_{\nu}\sim 3\mu_{R}$  on the mass parameters should be imposed.
If, for example, we adopt that the $\mu$ parameters originate from non-perturbative effects,
we expect that $\langle H\bar H\rangle\sim 10^{-2}M_S$, or $\langle H\rangle\sim 10^{-1}M_S$, in accordance
with our analysis.

From the D-flatness conditions we have:
\bea
&&\sum_{j}Q_j^{U(1)_4} |\Phi_j|^2~=~|\langle H \rangle|^2-|\langle \bar H \rangle|^2~=~0\\
&&\sum_{j}Q_j^{U(1)_{2L}} |\Phi_j|^2~=~
3\times 2|\langle \nu \rangle|^2+2\times (-2) |\langle \bar \nu \rangle|^2
 ~=~-\frac{ 6} {192\pi^2}\,g_s^2M_{S}^2\\
&&\sum_{j}Q_j^{U(1)_{2R}} |\Phi_j|^2~=~|\langle H \rangle|^2-|\langle \bar H \rangle|^2+2|\langle \D_R \rangle|^2~=~-\frac{- 6} {192\pi^2}\,g_s^2M_{S}^{2}
\eea
Solving the above system we find:
\bea
&&|\langle H \rangle|=|\langle \bar H \rangle|\\
&&|\langle \nu \rangle|=
\frac{2 \sqrt{2}}{3 {\cal M}} M_{S} \left(\mu _H-2 \mu _{\nu }\right)\\
&& |\langle \bar \nu \rangle|= \frac{\cal M}{2 \sqrt{2}}\\
&&|\langle \D_R \rangle|\sim \frac{g_s} {8\pi}\,M_{S}
\eea
where ${\cal M}=
{\sqrt{\frac{g_s^2 M_S^2}{32 \pi ^2}-\sqrt{\frac{g_s^4 M_S^4}{1024 \pi
   ^4}+\frac{32 M_S^2}{3} \left(\mu _H-2 \mu _{\nu }\right){}^2}}}$.

We notice that the $D$-flatness conditions determine  the scales of
$\langle\nu \rangle,\langle\bar\nu \rangle$ and $\langle\Delta_R \rangle$ vevs,
correlating them with the scale $M= \frac{g_s} {8\pi}\,M_{S}$, but
leave  the vevs for the $SU(4)$ breaking Higgs fields
$H,\bar H$. completely unspecified. Thus, we are  free to choose  the GUT scale
$M_{GUT}\sim \langle H\rangle$ appropriately to reconcile  other phenomenological
requirements, such as  fermion mass hierarchy and other aspects
related to renormalization group analysis of various low energy measured
quantities.

\subsection{Second example}

Here, we collect the  ${\cal W}_{3,4}$ superpotential terms of the second model case 1.
\bea
 {\cal W}_3 &=&\bar F_R F_L h' + (\bar F_R' + 2 \bar H') \D_R H + \bar h \bar h' \nu +  \bar h \bar h' \D_L +  \bar h \D_R h'\nn\\
{\cal W}_4 &=&
\bar h \bar F_R F_L \nu +  \bar h \bar F_R \D_L F_L +  (\bar F_R' + 2 \bar H') \bar h' F_L \nu +
(\bar F_R' + 2 \bar H') \bar h' \D_L F_L +  (\bar F_R' + 2 \bar H') \D_R F_L h' \nn\\
&&+  \bar F_R^2 H^2 +
\bar \D_L \bar F_R \D_L H +  \bar F_R \bar h' H h' +  \bar \D_R \bar F_R \D_R H +  \bar \D_R
\bar h'^2 \nu +  \bar h^2 \D_R \nu +  \bar \D_L^2 \D_L^2 \nn\\
&&+  \bar \D_L \bar h' \D_L
h' +  \bar \D_L \bar \D_R \D_L \D_R +  \bar \D_R \bar h'^2 \D_L +  \bar h^2 \D_L \D_R +  \bar \D_L
\D_R h'^2 +  \bar h'^2 h'^2 \nn\\
&&+  \bar \D_R \bar h' \D_R h' +  \bar \D_R^2 \D_R^2
\eea
Case 2 can be easily extracted from the above terms after suitable field redefinitions.

\subsection{Third example}

Here we give a third case to be worked out similarly
to the two examples given in the main text.
There are $two$ models with identical massless spectrum:
%
%
\bea
\begin{array}{llrcl}\nonumber
& Gauge~Group & Chirality  & ~~~~~~~~ & Spectrum\\
& U4\times U2\times U2 \\
 5 \times & (  V , \bar V ,0 ) & 3 & \to &  4  F_L +\bar F_L \\
 4 \times & (  V ,0 , \bar V) & 0 & \to &  2 H + 2 \bar  H\\
 3 \times & (  V ,0 ,V ) & -3 & \to &  3 \bar F_R' \\
 3 \times & (  A ,0 ,0 ) & 3 & \to & 3 D_6 \\
 8 \times & (  S ,0 ,0 ) & 0 & \to &  4 S_{10} +  4 \bar S_{10} \\
 5 \times & (  0 ,0 ,S ) & 3 & \to & 4 \D_R +  \bar \D_R \\
 6 \times & (  0 ,0 ,A ) & 0 & \to &  3 \f  + 3 \bar \f  \\
 6 \times & (  0 , \bar V , \bar V) & 0 & \to &  3 h + 3 \bar h \\
 3 \times & (  0 , \bar A ,0 ) & 3 & \to &  3 \bar \n \\
 3 \times & (  0 , \bar V ,V ) & -3 & \to &  3 h' \\
 4 \times & (  Adj,0 ,0 ) & 0 & \to &  0 \\
 1 \times & (  0 ,Adj,0 ) & 0 & \to &  0 \\
 4 \times & (  0 ,0 ,Adj) & 0 & \to &  0 \\
\end{array}\eea
but different hypercharge embedding.

These models, have the same internal sector that consists of a tensor product of six copies of ${\cal N}=2$ superconformal minimal models with levels $k_i=\{1,1,1,1,7,16\}$, and
they contain a single orientifold plane.

After the breaking:
In the first, the hypercharge is embedded as:
%
$Y={1\over 6}A_3+{1\over 2}A_1+{1\over 2}A_1'$
%
%
and in the other as:
$Y=-{1\over 3}A_3+{1\over 2}A_2$.
Both models have two extra massless U(1)'s:  $B_1={1\over 2}A_2+{1\over 2}A_1$,
$B_2={1\over 2}A_2-A_1'$.

This model has: $n=1,~m=0,~\bar n= 2,~\bar m= 0$. The trilinear couplings are:
\bea
{\cal W}_3 &=&
\bar h \phi h '  +  \bar  \n  h h '  +  \bar H F_L  h '  +
  \bar h D_R  h '  +  \bar F_R  '  \phi H \nn\\
  &&+  \bar F_R  '  \bar H D_6 +
  \bar F_R  '  F_L  h +  \bar F_R  '  \bar H S_{10}  +  \bar F_R  '  D_R  H\\
%
%
{\cal W}_4 &=&
\bar F_L \bar h \bar H D_6 +  \bar \Delta_R \bar H^2 D_6 +  \bar \phi \bar H^2 D_6 +
 \bar F_L^2 \bar \nu D_6 +  \bar F_R '^2 D_6 \Delta_R \nn\\ &&+  \bar \Delta_R^2 \Delta_R^2 +
 \bar F_R '^2 D_6 \phi +  \bar \phi^2 \phi^2 +  \bar \Delta_R \bar F_L \Delta_R F_L +
 \bar \phi \bar F_L \phi F_L \nn\\ &&+  \bar F_L^2 F_L^2 +  \bar \Delta_R \bar h \Delta_R h +  \bar \phi \bar h \phi h +
 \bar F_L \bar h F_L h +  \bar \Delta_R \bar H F_L h \nn\\ &&+  \bar \phi \bar H F_L h +  \bar h^2 h^2 +
 \bar \Delta_R \bar \nu h^2 +  \bar \phi \bar \nu h^2 +  \bar F_L \bar h \Delta_R H \nn\\ &&+  \bar \Delta_R \bar H \Delta_R H +
 \bar F_L \bar h \phi H +  \bar \phi \bar H \phi H +  \bar F_L \bar H F_L H +  \bar h \bar H h H \nn\\ &&+
 \bar F_L \bar \nu h H +  \bar S_{10} F_L h H +  \bar H^2 H^2 +  \bar S_{10} \Delta_R H^2 +
 \bar S_{10} \phi H^2 \nn\\ &&+  \bar F_R ' \Delta_R F_L h ' +  \bar F_R ' \phi F_L h ' +
 \bar \nu \Delta_R h '^2 +  \bar \nu \phi h '^2 +  \bar F_L \bar h \bar H S_{10} \nn\\ &&+
 \bar \Delta_R \bar H^2 S_{10} +  \bar \phi \bar H^2 S_{10} +  \bar F_L^2 \bar \nu S_{10} +
 \bar F_R '^2 \Delta_R S_{10} +  \bar \Delta_R \bar S_{10} \Delta_R S_{10} +  \bar S_{10}^2 S_{10}^2 \nn\\ &&+  \bar F_R '^2 \phi S_{10} +
 \bar \phi \bar S_{10} \phi S_{10} +  \bar F_L \bar S_{10} F_L S_{10} +  \bar h \bar S_{10} h S_{10} +
 \bar H \bar S_{10} H S_{10}
\eea

\section{Rest of Pati-Salam models at Gepner points}

In this appendix, we present several other consistent Pati-Salam Gepner vacua with or without additional branes \cite{Anastasopoulos:2006da}.

\subsection{Without Hidden Sector}

\begin{itemize}

\item Three models with internal sector that consists of a tensor product of five copies of ${\cal N}=2$ superconformal minimal models with levels $k_i=\{1,1,1,7,16\}$, a single orientifold plane
and spectra:
%
%
%
%
\bea
\begin{array}{llrcl}\nonumber
& Gauge~Group & Chirality  & ~~~~~~~~ & Spectrum\\
& U4\times U2\times U2 \\
 3 \times & (  A ,0 ,0 ) & 3 & \to & 3 D_6 \\
 3 \times & (  0 ,A ,0 ) & 3 & \to &  3 \n \\
 5 \times & (  V ,V ,0 ) & 3 & \to &  \bar F_L'  + 4  F_L'  \\
 5 \times & (  0 ,0 ,S ) & 3 & \to &  \bar \D_R  + 4 \D_R  \\
 3 \times & (  0 ,V ,V ) & -3 & \to &  3 \bar h \\
 3 \times & (  V ,0 ,V ) & -3 & \to &  3 \bar F_R' \\
 4 \times & (  Adj,0 ,0 ) & 0 & \to &  0 \\
 1 \times & (  0 ,Adj,0 ) & 0 & \to &  0 \\
 4 \times & (  0 ,0 ,Adj) & 0 & \to &  0 \\
 6 \times & (  0 ,0 ,A ) & 0 & \to &  3 \bar \f  + 3 \f  \\
 8 \times & (  S ,0 ,0 ) & 0 & \to &  4 \bar S_{10} + 4 S_{10} \\
 6 \times & (  0 ,V , \bar V) & 0 & \to &  3 \bar h' + 3 h' \\
 4 \times & (  V ,0 , \bar V) & 0 & \to &  2\bar H + 2H
\end{array}\eea
but different hypercharge embedding after the breaking:
In the first, the hypercharge is embedded as:
%
$Y={1\over 6}A_3+{1\over 2}A_1+{1\over 2}A_1'$
%
%
and in the other as:
$Y=-{1\over 3}A_3+{1\over 2}A_2$.
Both models have two extra massless U(1)'s:  $B_1={1\over 2}A_2+{1\over 2}A_1$,
$B_2={1\over 2}A_2-A_1'$.

\item A model with internal sector that consists of a tensor product of six copies of ${\cal N}=2$ superconformal minimal models with levels $k_i=\{1,1,1,1,7,16\}$, a single orientifold plane
and spectrum:
%
%
\bea
\begin{array}{llrcl}\nonumber
& Gauge~Group & Chirality  & ~~~~~~~~ & Spectrum\\
& U4\times U2\times U2 \\
 9 \times & (  0 ,0 ,S ) & 3 & \to &  3 \bar \D_R  + 6 \D_R  \\
 3 \times & (  A ,0 ,0 ) & 3 & \to &  3 D_6 \\
 3 \times & (  0 ,A ,0 ) & 3 & \to &  3 \n \\
 3 \times & (  V ,0 ,V ) & -3 & \to &  3 \bar F_R' \\
 3 \times & (  0 ,V ,V ) & -3 & \to &  3 \bar h \\
 5 \times & (  V ,V ,0 ) & 3 & \to &  \bar F_L'  + 4  F_L'  \\
 6 \times & (  Adj,0 ,0 ) & 0 & \to &  0 \\
 4 \times & (  S ,0 ,0 ) & 0 & \to &  2 \bar S_{10} + 2 S_{10} \\
 8 \times & (  0 ,S ,0 ) & 0 & \to &  4 \bar \Delta_L + 4 \Delta_L \\
 3 \times & (  0 ,Adj,0 ) & 0 & \to &  0 \\
 2 \times & (  0 ,0 ,Adj) & 0 & \to &  0 \\
 8 \times & (  V ,0 , \bar V) & 0 & \to &  4 \bar H + 4 H \\
 6 \times & (  0 ,V , \bar V) & 0 & \to &  3 \bar h' + 3 h'
\end{array}\eea
and hypercharge embedding after breaking:
%
$Y={1\over 6}A_3+{1\over 2}A_1+{1\over 2}A_1'$.
This model contains an extra massless U(1)'s  $B={1\over 2}A_1+A_1'$.
%
%
%

\item A model with internal sector that consists of a tensor product of five copies of ${\cal N}=2$ superconformal minimal models with levels $k_i=\{1,1,7,7,7\}$, a single orientifold plane
and spectrum:
%
%
\bea
\begin{array}{llrcl}\nonumber
& Gauge~Group & Chirality  & ~~~~~~~~ & Spectrum\\
& U4\times U2\times U2 \\
 6 \times & (  A ,0 ,0 ) & 6 & \to &  6 D_6 \\
 6 \times & (  0 ,A ,0 ) & 6 & \to &  6 \n \\
 3 \times & (  0 ,0 ,A ) & 3 & \to &  3 \f \\
 5 \times & (  0 ,0 ,S ) & 3 & \to &  \bar \D_R  + 4 \D_R  \\
 5 \times & (  V ,V ,0 ) & 3 & \to &  \bar F_L'  + 4  F_L'  \\
 3 \times & (  V ,0 ,V ) & -3 & \to &  3 \bar F_R' \\
 7 \times & (  Adj,0 ,0 ) & 0 & \to &  0 \\
 4 \times & (  0 ,Adj,0 ) & 0 & \to &  0 \\
 7 \times & (  0 ,0 ,Adj) & 0 & \to &  0 \\
 2 \times & (  S ,0 ,0 ) & 0 & \to &  \bar S_{10} + S_{10} \\
12 \times & (  V , \bar V,0 ) & 0 & \to &  6 \bar F_L + 6 F_L\\
 4 \times & (  V ,0 , \bar V) & 0 & \to & 2\bar H + 2H
\end{array}\eea
and hypercharge embedding after breaking:
%
$Y={1\over 6}A_3+{1\over 2}A_1+{1\over 2}A_1'$.
All the remaining abelian factors are massive due to anomalies.

\end{itemize}

\subsection{With Hidden Sector}

We have  several vacua where the massless spectrum consists of the Pati Salam branes (a stack of 4 and two stacks of 2 branes)
plus additional branes. Here we present  vacua with at most three stacks of additional branes:

\begin{itemize}

\item A model with internal sector that consists of a tensor product of five copies of ${\cal N}=2$ superconformal minimal models with levels $k_i=\{1,4,4,4,4\}$, a single orientifold plane
and spectrum:
%
%
\bea
\begin{array}{llrcl}\nonumber
& Gauge~Group & Chirality  & ~~~~~~~~ & Spectrum\\
&      U4\times U2\times U2 \times O4\\
      5 \times & ( V ,V ,0 ;0 ) &   3       & \to & 4 F_L +  \bar F_L\\
      3 \times & ( V ,0 , \bar V;0 ) &   -1      & \to & F_R, H, \bar H\\
      2 \times & ( V ,0 ,V ;0 ) &   -2      & \to & 2F_R'\\
      2 \times & ( A ,0 ,0 ;0 ) &   0       & \to &  D_6  +  \bar D_6 \\
     10 \times & ( 0 ,S ,0;0 ) &   0       & \to & 5\Delta_L + 5\bar \Delta_L \\
      1 \times & ( 0 ,0 ,A ;0 ) &   1       & \to & \f \\
      3 \times & ( 0 ,V ,V ;0 ) &   -3      & \to & 3 h\\
      5 \times & ( 0 ,V , \bar V;0 ) &   -3      & \to & 4h' +  \bar h'\\
      1 \times & ( 0 ,0 ,S ;0 ) &   1       & \to & \Delta_R\\
      1 \times & ( Adj,0 ,0 ;0 ) &   0\\
      4 \times & ( 0 ,Adj,0 ;0 ) &   0\\
      1 \times & ( 0 ,0 ,Adj;0 ) &   0\\
      6 \times & ( V ,0 ,0 ;V ) &   0\\
      2 \times & ( 0 ,0 ,V ;V ) &   0\\
      2 \times & ( 0 ,V ,0 ;V ) &   0\\
      2 \times & ( 0 ,0 ,0 ;S ) &   0\\
      4 \times & ( 0 ,0 ,0 ;A ) &   0\\
      \end{array}\eea
%
%
%
and hypercharge after breaking:
$Y={1\over 6}A_3+{1\over 2}A_1+{1\over 2}A_1'$.
It also contains another massless U(1): $B={1\over 3}A_2+A_1$.

\item A model with internal sector that consists of a tensor product of five copies of ${\cal N}=2$ superconformal minimal models with levels $k_i=\{1,4,4,4,4\}$, a single orientifold plane
and spectrum:
%
%
\bea
\begin{array}{llrcl}\nonumber
& Gauge~Group & Chirality  & ~~~~~~~~ & Spectrum\\
&      U4\times U2\times U2 \times U1 \times U2\\
      1 \times & ( V , \bar V,0 ;0 ,0 ) &   1      & \to & F_L\\
      2 \times & ( V ,V ,0 ;0 ,0 ) &   2      & \to & 2 F_L'\\
      1 \times & ( V ,0 , \bar V;0 ,0 ) &   -1     & \to & F_R\\
      2 \times & ( V ,0 ,V ;0 ,0 ) &   -2     & \to & 2 F_R'\\
      2 \times & ( 0 ,V , \bar V;0 ,0 ) &   -2     & \to & 2 \bar h'\\
      1 \times & ( 0 ,V ,V ;0 ,0 ) &   1      & \to & 2 h\\
      1 \times & ( 0 ,A ,0 ;0 ,0 ) &   1      & \to & \n's\\
      2 \times & ( 0 ,0 ,A ;0 ,0 ) &   -2     & \to & 2\bar \f\\
      3 \times & ( 0 ,0 ,S ;0 ,0 ) &   -1     & \to & \Delta_R+  2\bar \Delta_R\\
      2 \times & ( 0 ,V ,0 ;V ,0 ) &   0\\
      2 \times & ( V ,0 ,0 ; \bar V,0 ) &   0\\
      2 \times & ( V ,0 ,0 ;V ,0 ) &   0\\
      2 \times & ( 0 ,V ,0 ; \bar V,0 ) &   0\\
      1 \times & ( 0 ,0 ,Adj;0 ,0 ) &   0\\
     11 \times & ( 0 ,0 ,0 ;Adj,0 ) &   0\\
      6 \times & ( 0 ,0 ,V ; \bar V,0 ) &   0\\
      2 \times & ( 0 ,0 ,V ;V ,0 ) &   0\\
     12 \times & ( 0 ,0 ,0 ;A ,0 ) &   0\\
      8 \times & ( 0 ,0 ,0 ;S ,0 ) &   0\\
      2 \times & ( 0 ,0 ,0 ;V , \bar V) &   0\\
      2 \times & ( 0 ,0 ,0 ;V ,V ) &   0\\
      \end{array}\eea
and hypercharge embedding after breaking:
$Y={1\over 6}A_3+{1\over 2}A_1+{1\over 2}A_1'$.
All remaining abelian factors are massive due to anomalies.

\item A model with internal sector that consists of a tensor product of five copies of ${\cal N}=2$ superconformal minimal models with levels $k_i=\{1,1,2,14,46\}$, a single orientifold plane
and spectrum:
%
%
\bea
\begin{array}{llrcl}\nonumber
& Gauge~Group & Chirality  & ~~~~~~~~ & Spectrum\\
&      U4\times U2\times U2 \times U2 \times U2\\
      7 \times & ( 0 ,0 ,S ;0 ,0 ) &   3     & \to & 5 \Delta_R+ 2 \bar \Delta_R\\
      3 \times & ( V ,0 ,V ;0 ,0 ) &   -3    & \to & 3 F_R\\
      3 \times & ( 0 ,A ,0 ;0 ,0 ) &   3     & \to & 3 \n's\\
      3 \times & ( V ,V ,0 ;0 ,0 ) &   3     & \to & 3 F_L\\
      3 \times & ( 0 ,V ,V ;0 ,0 ) &   -3    & \to & 3 h\\
      8 \times & ( 0 ,S ,0 ;0 ,0 ) &   0     & \to & 4 \Delta_L + 4 \bar \Delta_L \\
      4 \times & ( V ,0 , \bar V;0 ,0 ) &   0     & \to & 2 H + 2 \bar H\\
      4 \times & ( 0 ,V , \bar V;0 ,0 ) &   0     & \to & 2h' + 2 \bar h'\\
      1 \times & ( Adj,0 ,0 ;0 ,0 ) &   0\\
      2 \times & ( 0 ,Adj,0 ;0 ,0 ) &   0\\
      3 \times & ( 0 ,0 ,Adj;0 ,0 ) &   0\\
      2 \times & ( 0 ,0 ,0 ;0 ,Adj) &   0\\
      6 \times & ( 0 ,V ,0 ;0 ,V ) &   0\\
      2 \times & ( 0 ,0 ,V ; \bar V,0 ) &   0\\
      \end{array}\eea
%
%
%
and hypercharge embedding after breaking:
$Y={1\over 6}A_3+{1\over 2}A_1+{1\over 2}A_1'$
All remaining abelian factors are massive due to anomalies.

\item A model with internal sector that consists of a tensor product of four copies of ${\cal N}=2$ superconformal minimal models with levels $k_i=\{2,10,10,10\}$, a single orientifold plane
and spectrum:
%
%
\bea
\begin{array}{llrcl}\nonumber
& Gauge~Group & Chirality  & ~~~~~~~~ & Spectrum\\
&      U4\times U2\times U2 \times U2\\
      3 \times & ( 0 ,0 ,S ;0 ) &   1     & \to & 2 \Delta_R+  \bar \Delta_R\\
      4 \times & ( V ,V ,0 ;0 ) &   2     & \to & 3 F_L'+  \bar F_L'\\
      3 \times & ( V ,0 , \bar V;0 ) &   -1    & \to & F_R, H,\bar H\\
      3 \times & ( 0 ,V ,V ;0 ) &   -1    & \to & 2 h+  \bar h\\
      3 \times & ( V , \bar V,0;0 ) &   1     & \to & 2 F_L+  \bar F_L\\
      2 \times & ( V ,0 ,V ;0 ) &   -2    & \to & 2 F_R'\\
      1 \times & ( 0 ,A ,0 ;0 ) &   1     & \to & \n's\\
      6 \times & ( 0 ,S ,0 ;0 ) &   0     & \to & 3 \Delta_L + 3 \bar \Delta_L \\
      4 \times & ( 0 ,V , \bar V;0 ) &   0     & \to & 2 h'+ 2 \bar h'\\
      4 \times & ( S ,0 ,0 ;0 ) &   0     & \to & 2 S_{10}  + 2 \bar S_{10}\\
      2 \times & ( A ,0 ,0 ;0 ) &   0     & \to & 1 D_6  +  \bar D_6  \\
      2 \times & ( Adj,0 ,0 ;0 ) &   0\\
      2 \times & ( 0 ,Adj,0 ;0 ) &   0\\
      1 \times & ( 0 ,0 ,Adj;0 ) &   0\\
      2 \times & ( 0 ,V ,0 ; \bar V) &   0
      \end{array}\eea
%
%
%
and hypercharge embedding after breaking:
$Y={1\over 6}A_3+{1\over 2}A_1+{1\over 2}A_1'$.
All the remaining abelian factors are massive due to anomalies.

\item A model with internal sector that consists of a tensor product of five copies of ${\cal N}=2$ superconformal minimal models with levels $k_i=\{1,4,4,4,4\}$, a single orientifold plane
and spectrum:
%
%
%
\bea
\begin{array}{llrcl}\nonumber
& Gauge~Group & Chirality  & ~~~~~~~~ & Spectrum\\
&      U4\times U2\times U2 \times U2 \times Sp6\\
      2 \times & ( 0 ,A ,0 ;0 ,0 ) &   2     & \to & 2 \n's\\
      2 \times & ( 0 ,0 ,A ;0 ,0 ) &   -2    & \to & 2 \bar \Delta_R\\
      3 \times & ( V ,V ,0 ;0 ,0 ) &   3     & \to & 3 F_L\\
      3 \times & ( V ,0 ,V ;0 ,0 ) &   -3    & \to & 3 F_R\\
      2 \times & ( S ,0 ,0 ;0 ,0 ) &   0     & \to & S_{10}  + \bar S_{10}\\
      2 \times & ( 0 ,V ,V ;0 ,0 ) &   0     & \to & h +  \bar h\\
      4 \times & ( 0 ,V , \bar V;0 ,0 ) &   -4    & \to & 2h' + 2 \bar h'\\
      2 \times & ( 0 ,V ,0 ;0 ,V ) &   0\\
      2 \times & ( 0 ,0 ,V ;0 ,V ) &   0\\
      7 \times & ( 0 ,0 ,0 ;A ,0 ) &   3\\
      3 \times & ( 0 ,0 ,0 ;V ,V ) &   -1\\
      2 \times & ( 0 ,0 ,0 ;S ,0 ) &   2\\
      2 \times & ( 0 ,0 ,0 ;Adj,0 ) &   0\\
      2 \times & ( 0 ,0 ,0 ;0 ,A ) &   0\\
      \end{array}\eea
%
%
%
and hypercharge embedding after breaking:
$Y={1\over 6}A_3+{1\over 2}A_1+{1\over 2}A_1'$.
All the remaining abelian factors are massive due to anomalies.

\item A model with internal sector that consists of a tensor product of four copies of ${\cal N}=2$ superconformal minimal models with levels $k_i=\{3,8,8,8\}$, a single orientifold plane
and spectrum:
%
%
\bea
\begin{array}{llrcl}\nonumber
& Gauge~Group & Chirality  & ~~~~~~~~ & Spectrum\\
&      U4\times U2\times U2 \times U4 \times U2 \times U4\\
      2 \times & ( 0 ,A ,0 ;0 ,0 ,0 ) &   2    & \to & 2 \n's\\
      2 \times & ( 0 ,0 ,A ;0 ,0 ,0 ) &   -2   & \to & 2 \f \\
      2 \times & ( V ,0 ,V ;0 ,0 ,0 ) &   -2   & \to & 2 F_R\\
      2 \times & ( V ,V ,0 ;0 ,0 ,0 ) &   2    & \to & 2 F_L\\
      1 \times & ( V , \bar V,0 ;0 ,0 ,0 ) &   1    & \to & F_L'\\
      1 \times & ( V ,0 , \bar V;0 ,0 ,0 ) &   -1   & \to & F_R'\\
      2 \times & ( A ,0 ,0 ;0 ,0 ,0 ) &   0    & \to & D_6  +  \bar D_6 \\
      2 \times & ( 0 ,V , \bar V;0 ,0 ,0 ) &   0    & \to & h' +  \bar h'\\
      2 \times & ( 0 ,V ,V ;0 ,0 ,0 ) &   0    & \to & h +  \bar h\\
      1 \times & ( Adj,0 ,0 ;0 ,0 ,0 ) &   0\\
      2 \times & ( V ,0 ,0 ; \bar V,0 ,0 ) &   0\\
      2 \times & ( V ,0 ,0 ;0 ,V ,0 ) &   0\\
      1 \times & ( 0 ,0 ,0 ;0 ,Adj,0 ) &   0\\
      2 \times & ( 0 ,0 ,0 ;0 ,0 ,Adj) &   0\\
      \end{array}\eea
%
%
%
and hypercharge embedding after breaking:
$Y={1\over 6}A_3+{1\over 2}A_1+{1\over 2}A_1'$.
All the remaining abelian factors are massive due to anomalies.

\item A model with internal sector that consists of a tensor product of five copies of ${\cal N}=2$ superconformal minimal models with levels $k_i=\{1,1,1,7,16\}$, a single orientifold plane
and spectrum:
%
%
%
\bea
\begin{array}{llrcl}\nonumber
& Gauge~Group & Chirality  & ~~~~~~~~ & Spectrum\\
&      U4\times U2\times U2\times U4 \\
      3 \times & ( A ,0 ,0 ;0 ) &   3    & \to & 3 D_6 \\
      6 \times & ( 0 ,A ,0 ;0 ) &   6    & \to & 6 \n's\\
      3 \times & ( 0 ,0 ,A ;0 ) &   3    & \to & 3 \f \\
      5 \times & ( 0 ,0 ,S ;0 ) &   3    & \to & 4  \Delta_R+  \bar \Delta_R\\
      5 \times & ( V ,V ,0 ;0 ) &   3    & \to & 4 F_L+  \bar F_L\\
      3 \times & ( V ,0 ,V ;0 ) &   -3   & \to & 3 F_R\\
     12 \times & ( V , \bar V,0 ;0 ) &   0    & \to & 6 F_L'+ 6\bar F_L'\\
      6 \times & ( V ,0 , \bar V;0 ) &   0    & \to & 3 H+ 3 \bar H \\
      1 \times & ( Adj,0 ,0 ;0 ) &   0\\
      4 \times & ( 0 ,Adj,0 ;0 ) &   0\\
      7 \times & ( 0 ,0 ,Adj;0 ) &   0\\
      2 \times & ( 0 ,0 ,V ; \bar V) &   0\\
      6 \times & ( 0 ,0 ,V ;V ) &   0\\
      2 \times & ( V ,0 ,0 ; \bar V) &   0\\
      6 \times & ( V ,0 ,0 ;V ) &   0\\
      3 \times & ( 0 ,0 ,0 ;A ) &   -3\\
      1 \times & ( 0 ,0 ,0 ;Adj) &   0\\
      \end{array}\eea
%
%
%
and hypercharge embedding after breaking:
$Y={1\over 6}A_3+{1\over 2}A_1+{1\over 2}A_1'$.
It also contains another massless U(1): $B={1\over 2}A_2+{1\over 2}A_1+ 2A_1'$.

\end{itemize}

\end{document}